\documentclass[aps,prx,twocolumn,superscriptaddress]{revtex4-2}

\usepackage{amsmath}
\usepackage{amssymb}
\usepackage{amsthm}
\usepackage{amsfonts}
\usepackage{listings}
\usepackage{physics}
\usepackage{enumerate}
\usepackage{latexsym}
\usepackage{color}
\usepackage{bm}
\usepackage{hyperref}
\hypersetup{
 pdfnewwindow=true, colorlinks=true,
 linkcolor=blue, anchorcolor=blue,
 citecolor=blue, filecolor=blue,
 menucolor=blue, urlcolor=blue}

\usepackage{psfrag}

\usepackage{bm}
\usepackage{graphicx}
\usepackage{subfigure}

\usepackage{multirow}
\usepackage{array}

\RequirePackage[normalem]{ulem} 
\RequirePackage{color}\definecolor{RED}{rgb}{1,0,0}\definecolor{BLUE}{rgb}{0,0,1}

\newcommand{\ovl}[1]{\overline{#1}}

\newcommand{\bx}{{\vb* x}}
\newcommand{\bE}{{\vb* E}}
\newcommand{\bB}{{\vb* B}}

\newcommand{\bc}{{\vb* c}}

\newcommand{\bS}{{\vb* S}}
\newcommand{\bk}{{\vb* k}}

\newcommand{\bP}{{\vb* P}}
\newcommand{\bA}{{\vb* A}}
\newcommand{\bl}{{\vb* l}}

\newcommand{\pref}[1]{(\ref{#1})}

\def\ie{{\it i.e.},\ }
\def\eg{{\it e.g.}\ }

\input{epsf}

\newcommand{\nc}{\newcommand}
\nc{\webirvsp}{\href{https://github.com/zjwang11/irvsp}{\texttt{IRVSP}} }
\nc{\webirtb}{\href{https://github.com/zjwang11/irvsp}{\texttt{ir2tb}} }
\nc{\webirpw}{\href{https://github.com/zjwang11/ir2pw}{\texttt{ir2ph}} }
\nc{\webchecktopmat}{\href{https://www.cryst.ehu.es/cryst/checktopologicalmagmat}{\texttt{Check Topological Mat}}}
\nc{\webposabr}{\href{https://github.com/zjwang11/UnconvMat/blob/master/src_pos2aBR.tar.gz}{\texttt{POS2ABR}} }
\nc{\webUnconvMat}{\href{http://tm.iphy.ac.cn/UnconvMat.html}{\texttt{UnconvMat}} }
\nc{\online}{\href{http://tm.iphy.ac.cn/UnconvMat.html}{online}}

\begin{document}

\tolerance 10000

%\draft

\title{Nonsymmorphic Time-Reversal-Doubled Axion Insulators}

% in English titles articles and words like to, on, at etc are always spelled with small letters

\author{Yue Xie}
\affiliation{Beijing National Laboratory for Condensed Matter Physics,
and Institute of Physics, Chinese Academy of Sciences, Beijing 100190, China}
\affiliation{School of Physical Sciences, University of Chinese Academy of Sciences, Beijing 100049, China}

\author{Haohao Sheng}
\affiliation{Beijing National Laboratory for Condensed Matter Physics,
and Institute of Physics, Chinese Academy of Sciences, Beijing 100190, China}
\affiliation{School of Physical Sciences, University of Chinese Academy of Sciences, Beijing 100049, China}
\author{Quansheng~Wu}
\affiliation{Beijing National Laboratory for Condensed Matter Physics,
and Institute of Physics, Chinese Academy of Sciences, Beijing 100190, China}
\affiliation{Condensed Matter Physics Data Center, Chinese Academy of Sciences, Beijing 100190, China}

\author{Xi~Dai}
\affiliation{Department of Physics, Hong Kong University of Science and Technology, Hong Kong 999077, China}
\author{Zhong Fang}
\affiliation{Beijing National Laboratory for Condensed Matter Physics,
and Institute of Physics, Chinese Academy of Sciences, Beijing 100190, China}
\affiliation{Condensed Matter Physics Data Center, Chinese Academy of Sciences, Beijing 100190, China}

\author{Hongming~Weng}
\email{hmweng@iphy.ac.cn}
\affiliation{Beijing National Laboratory for Condensed Matter Physics,
and Institute of Physics, Chinese Academy of Sciences, Beijing 100190, China}
\affiliation{Condensed Matter Physics Data Center, Chinese Academy of Sciences, Beijing 100190, China}
\author{Zhijun~Wang}
\email{wzj@iphy.ac.cn}
\affiliation{Beijing National Laboratory for Condensed Matter Physics,
and Institute of Physics, Chinese Academy of Sciences, Beijing 100190, China}
\affiliation{Condensed Matter Physics Data Center, Chinese Academy of Sciences, Beijing 100190, China}

\date{\today}

\begin{abstract}
The axion insulator exhibits a topological magnetoelectric effect characterized by an axion angle $\theta=\pi$, while the time-reversal-doubled axion insulator ($T$-DAXI) can be viewed as two copies of an axion insulator related by time-reversal symmetry. In this work, we show that a nonsymmorphic topological crystalline insulator with glide (or screw) symmetry can host the nonsymmorphic $T$-DAXI phase, diagnosed by the nontrivial hourglass (screw) invariant $\delta_h=1$ ($\delta_s=1$).
In this phase, the partial axion angles ($\theta^\pm=\pi $) are quantized by the glide and/or screw symmetry, and the gapped surfaces realize half-quantized quantum spin Hall states.
Furthermore, we show that nonsymmorphic $T$-DAXIs also exhibit the spin-magnetoelectric effect. Namely, an external magnetic field can induce a topological nonlocal spin polarization in real space.
When the magnetic field is time-periodic, this leads to an alternating spin current detectable in experiment.
Using \emph{ab initio} calculations, we demonstrate that mixed bismuth monohalides Bi$_4$Br$_{x}$I$_{4-x}$ realize the nonsymmorphic $T$-DAXI.
%Finally, using \emph{ab initio} calculations and the spin-resolved nested Wilson loop method, we find that $\gamma $-Bi$_4$Br$_4$ is a nonsymmorphic topological crystalline insulator, being a $T$-DAXI in the $s_y$ spin resolution. 
Our findings not only reveal the symmetry-enforced $T$-DAXIs in nonsymmorphic topological crystalline insulators, but also demonstrate the spin-magnetoelectric effect as a detectable topological spin response in nonsymmorphic $T$-DAXIs.
\end{abstract}

\maketitle

%\paragraph*{Introduction.}
\section{Introduction}
In recent decades, topological phases of matter have attracted sustained interest in condensed matter physics and materials science~\cite{review-TI-2010,review-TI-2011,BHZ,Kane-Mele-1,Kane-Mele-2,3DTI,Guo2022,PhysRevX.11.021016}.
One of the intriguing properties characteristic of three-dimensional (3D) topological materials is the topological magnetoelectric effect arising from an effective action $S_\theta=\left(\theta e^2 / 4\pi^2\right)\int dtd^3\bx \bE\cdot \bB$, where $\theta$ is the axion angle defined over a period of $2\pi$~\cite{AXI-Qi-prb,dynamical-AXI-Qi-prb,Wilczek-AXI,review-JoAP,review-naturereviewphysics}. An insulator with $\theta=\pi$ quantized by symmetry that reverses odd times of space-time coordinates (\eg $I$, $T\tau$, $M$, $C_{2}T$) is termed a (magnetic) axion insulator (AXI), and exhibits a detectable quantized magnetoelectric effect~\cite{Essin-AXI,TME-2021,ACcurrent-2022,AXI-AFM-2022}.
The AXI can be constructed from weakly coupled alternating Chern-insulator layers with $C=\pm 1$~\cite{magnetic-tqc,axionangle-WCC-2020,statistical-AXI-Song}.
Materials such as MnBi$_2$Te$_4$~\cite{MBT-Zhang,MBT-Duan,exp-MBT,MBT-exp-QAH}, Eu$_3$In$_2$As$_4$~\cite{EuInAs} and others~\cite{Ta2Se8I,EuSn2P2-exp,EuSn2P2-cal,EuIn2As2-prl,Cr-BiTe,Cr-BiTe-scienceadvanced,GdPtBi-2010,GdPtBi-2014} have been theoretically proposed as candidates for AXIs. Moreover, an AXI features half of a quantum anomalous Hall (QAH) state on a surface due to the parity anomaly~\cite{AXI-Vanderbilt-2018,Olsen-AXI,PhysRevLett.129.096601}. In fact, the half-quantized surface conductivity is carried by the chiral hinge states~\cite{10.1093/nsr/nwad025}, which links AXIs to chiral higher-order topological insulators (HOTIs), or magnetic topological crystalline insulators (TCIs)~\cite{BBH-science,BBH,HOTI-2017-PRL,d-2-Song,HOTI-Schindler,HOTI-Ezawa,magnetic-tqc-Xu,magnetic-tqc-Gao}.

On the other hand, time-reversal ($T$-) invariant TCIs~\cite{TCI-2011,mirrorTCI,hourglass-fermion,rotation-anomaly,tqc-Wang,tqc-Fang,tqc-Wan} host gapless helical boundary states at terminations that preserve protecting crystalline symmetries~\cite{Wieder2018}. An example is the helical hinge states in $T$-invariant inversion TCIs with inversion index $Z_4=2$~\cite{LC-Song,Z4-PRX,Z4-nc,Z4-prb,HOTI-MoTe2-Wang}, also known as helical HOTIs~\cite{HOTI-2017-PRL,d-2-Song,HOTI-Schindler}.
Recent work~\cite{the-nc} has shown that for a chosen spin-resolved direction, an inversion TCI can realize one of the three refined spin-resolved topological phases: a 3D quantum spin Hall insulator (QSHI), a spin Weyl-fermion state, and notably a $T$-invariant doubled axion insulator ($T$-DAXI).
The $T$-DAXI can be viewed as two copies of a magnetic AXI related by time-reversal symmetry, with nontrivial $\pi$ partial axion angles quantized by inversion symmetry.
The discovery of $T$-DAXIs has expanded the possible venues for axion physics, \eg half-quantized 2D surface QSHI.
Especially, novel spin-magnetoelectric responses are introduced in those inversion-based HOTIs and/or $T$-DAXIs~\cite{spin-ME-2013,the-nc,spin-ME-2022}.
However, both $0$ and $\pi$ of the partial axion angles are allowed for the spin-resolved topology of an inversion TCI, and thus whether a $T$-DAXI can be realized depends on the details of the projected spin bands.
These facts raise three natural questions: (i) can $T$-DAXI phases arise with partial axion angles quantized by other crystalline symmetries, instead of inversion, (ii) can the $T$-DAXI phases be diagnosed by any topological invariants, and (iii) dose the spin-magnetoelectric effect still exist in the nonsymmorphic $T$-DAXI phases?

\begin{table*}[!t]
    \begin{ruledtabular}
    \caption{Spin-resolved topological classifications in spinful 3D nonsymmorphic TCIs. $\theta^+$ is the partial axion angle and $C^+$ is the partial Chern number in the upper spin subspace. Due to the time reversal symmetry, $\theta^+=-\theta^-$ and $C^+=-C^-$.}
        \centering
        \begin{tabular}{ccccc}
        %\hline
        Nonsymmorphic TCI  & \multicolumn{2}{c}{$g_y\equiv\left\{ M_{y} \,\middle|\, 0, 0, 0.5 \right\}$} & \multicolumn{2}{c}{$s_2\equiv\left\{ C_{2z} \,\middle|\, 0, 0, 0.5 \right\}$} \\
        \hline
        Spin direction & $y$ axis & $xz$ plane & $xy$ plane & $z$ axis \\
        \hline
        Symmetry in a subspace & $g_y$ & $g_yT$ & $s_2T$ & $s_2$ \\ 
        \hline
        Topological number & $\theta^{\pm}$ & $C^{\pm}$ & $\theta^{\pm}$ & $C^{\pm}$ \\
        \hline
        Classification and topology & $\mathbb{Z}_2$ ($T$-DAXI) & $\mathbb{Z}$ (3D QSHI) & $\mathbb{Z}_2$ ($T$-DAXI) & $\mathbb{Z}$ (3D QSHI) \\
        %\hline
        \end{tabular}    
        \label{table: sLC classification}
    \end{ruledtabular}
\end{table*}

In this work, we introduce the spin-resolved topological classifications for 3D $T$-invariant insulators with glide $g_y$ or screw $s_2$ symmetry, as shown in Table~\ref{table: sLC classification}. In the $s_y$ ($s_{xy}$) spin-resolved topology, we find that the $T$-DAXI phases are protected by the nonsymmorphic symmetry in TCIs with a nontrivial hourglass (screw) invariant $\delta_h=1$ ($\delta_s=1$). The partial axion angles of a nonsymmorphic T-DAXI defined in spin subspaces are quantized by the symmetry $g_y$ ($s_2T$). 
To illustrate the properties of the nonsymmorphic $T$-DAXI, we construct a glide TCI model as an example and explicitly reveal the following properties, including the quantized partial axion angles $\theta^{\pm}=\pi$, surface quantum anomalies with a surface half-quantized spin Chern number plateau $-C_{s,\text{top}}=C_{s,\text{bot}}=1/2$, and a topological spin-magnetoelectric effect that produces a detectable transport signature.
Finally, using \emph{ab initio} calculations, we predict that the mixed bismuth monohalides Bi$_4$BrI$_3$ and Bi$_4$Br$_3$I can realize the nonsymmorphic $T$-DAXIs, providing a promising platform for experimental detection.

\begin{figure}[t!]
\centering
\includegraphics[width=1\linewidth]{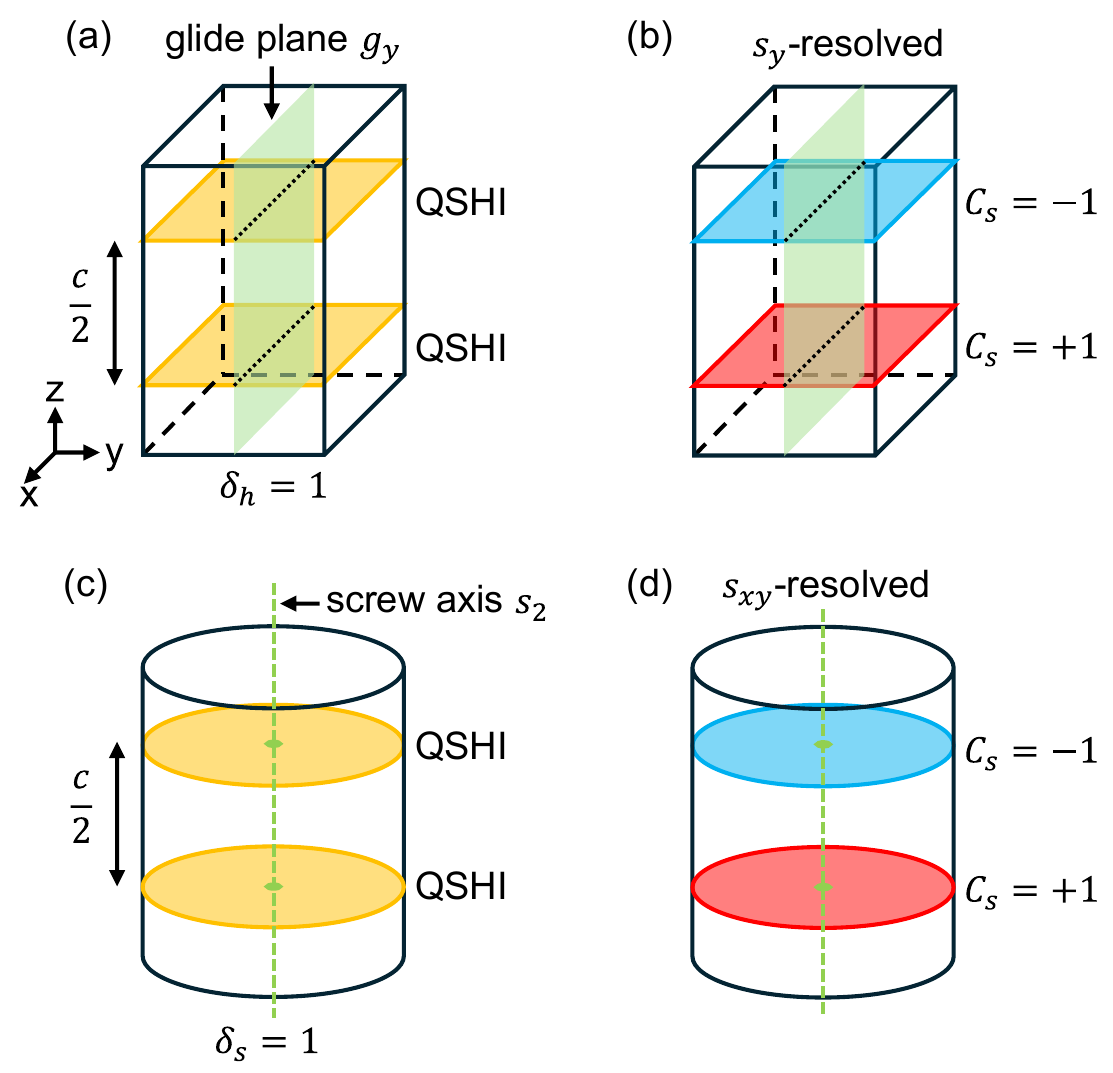}
\caption{(Color online) 
The layer constructions of the nonsymmorphic TCIs and symmetry-enforced spin-resolved phases of nonsymmorphic $T$-DAXIs.
(a) Layer construction of a gTCI with an hourglass invariant $\delta_h=1$, where two QSHIs are related by $g_y$.
(b) Symmetry-enforced spin-resolved topology of the glide $T$-DAXI in the $s_y$ spin direction for a gTCI of (a).
(c) Layer construction of a sTCI with a screw invariant $\delta_s=1$, where two QSHIs are related by $s_2$.
(d) Symmetry-enforced spin-resolved topology of the screw $T$-DAXI in the $s_{xy}$ spin direction for a sTCI of (c).
In (b) and (d), the two QSHIs carry opposite $s_y$ and $s_{xy}$ spin Chern numbers due to the glide and screw symmetry, respectively, both realizing a nonsymmorphic $T$-DAXI.
} \label{fig-LC}
\end{figure}

%\paragraph*{$T$-DAXIs in nonsymmorphic TCIs.}
\section{$T$-DAXIs in nonsymmorphic TCIs}\label{section: spin-resolved topology}
To investigate the spin-resolved topology, one has to separate the occupied-band space into two spin subspaces by projecting the $n$-directed spin operator $\hat{s}_n$~\cite{Prodan},
\begin{equation}\label{projectedspin}\begin{aligned}
    &S_{n}(\bk)=P(\bk)\hat{s}_nP(\bk),   \\
    &S_{n}\ket{\psi^{\pm}_{m\bk}}=\lambda^{\pm}_{m\bk}\ket{\psi^{\pm}_{m\bk}},
\end{aligned}\end{equation}
where $P(\bk)=\sum_{m\in\mathrm{occ}}\ket{\psi_{m\bk}}\bra{\psi_{m\bk}}$ projects onto the occupied-band space, and $n$ denotes the spin direction. When a spin gap exists between upper and lower spin bands, labeled by eigenvalues $\lambda^{+}_{m\bk}$ and $\lambda^{-}_{m\bk}$, the occupied bands can be separated into two sectors.
Thus in a QSHI, one can define partial Chern numbers $C^{\pm}$ for the two sectors, with time-reversal symmetry requiring $C^{+}=-C^{-}$.
Accordingly, non-zero spin Chern number $C_s\equiv(C^{+}-C^{-})/2$ indicates the existence of a topological contribution to the intrinsic spin Hall conductivity (SHC) $\sigma_s$~\cite{general-spinChern-PRL,general-spinChern-PRB}.
Provided a system is spin conserved, its intrinsic SHC will strictly equal to a quantized value, $\sigma_s= C_s e/2\pi$. When spin conservation is perturbatively broken (normally in real materials), the intrinsic SHC does not instantaneously vanish, but deviates from and lies close to such quantized value if the spin-nonconserving spin-orbit coupling (SOC) is relatively weak.
%In fact, the SHC $\sigma^{s,\hat{n}}\approx \tilde{\sigma}^{s,\hat{n}}$ if the spin-nonconserving spin-orbit coupling (SOC) is relatively weak.

%Hereafter, the nonsymmorphic TCIs are referred to as glide (screw) TCIs with the hourglass (screw) invariant $\delta_h=1$ ($\delta_s=1$) and four zero time-reversal $\mathbb{Z}_2$ indices $(\nu_0;\nu_1\nu_2\nu_3)=(0;000)$~\cite{LC-Song}.
A glide TCI (gTCI) can be constructed in real space by stacking QSHI layers in the $z$ direction, where two QSHIs per unit cell are related by $g_y$ and individually occupy the $(001;d_0)$ and $(001;d_0+c/2)$ planes [here $(hkl;d)$ denotes Miller indices $hkl$ and the distance $d$ to the origin], as shown in Fig.~\ref{fig-LC}(a).
Meanwhile, a screw TCI (sTCI) is constructed by simply replacing the glide plane with a two-fold screw axis. Such layer construction yields a nontrivial hourglass invariant $\delta_h=1$ or screw invariant $\delta_s=1$ with trivial strong and weak time-reversal $\mathbb{Z}_2$ indices $(\nu_0;\nu_1\nu_2\nu_3)=(0;000)$ for the gTCIs/sTCIs~\cite{LC-Song}, which we refer to as nonsymmorphic TCIs in this work.
For a gTCI, when the bands are separated by the eigenvalues of $P s_y P$, the $g_y$ symmetry is preserved in each spin sector and can quantize the partial axion angle. Based on spin-resolved bubble-equivalence analysis, the spin-resolved topology is $\mathbb Z_2$-classified (see Appendix~\ref{SM: bubble-equvalence}).
In Fig.~\ref{fig-LC}(b), since the two QSHIs are related by $g_y$, the $s_y$ spin Chern numbers of them are opposite, consistent with a nontrivial partial axion angle $\theta^\pm=\pi$ in both spin sectors [Fig.~\ref{fig-bubble}(b)].
Therefore, the system realizes the spin-resolved topological state of a $T$-DAXI.
For other spin-resolved directions, the $T$-DAXI phase is not allowed/protected because the partial axion angles are no longer quantized due to the lack of $g_y$ in each set of spin bands.
Additionally, when the bands are separated by $s_{xz}$ (we denote $s_{ab}$ for an arbitrary spin direction in the $ab$ plane), the partial Chern numbers for each QSHI layer are identical due to $g_yT$, and the spin-resolved topology is $\mathbb{Z}$-classified, corresponding to a spin-resolved 3D QSHI. 
Regarding an sTCI with $s_2$ symmetry, the spin-resolved topology for spin directions in the $xy$ plane yields $T$-DAXIs, whose partial axion angles are quantized by $s_2T$ symmetry, while the $s_z$ spin resolution produces an $s_2$-protected 3D QSHI.
In particular, unlike in inversion-symmetric TCIs that the spin-resolved topology ($T$-DAXI or 3D QSHI) depends on detailed spin-spectrum diagnosis~\cite{the-nc}, the $s_y$ ($s_{xy}$) spin-resolved topology of a nonsymmorphic $T$-DAXI is protected and indicated by the invariant of gTCIs (sTICs).

%\begin{figure}[tb!]
%\centering
%\includegraphics[width=8.0 cm,
%                 trim=0mm 0mm 0mm 0mm]{fig-hinge.pdf}
%\caption{(Color online) 
%(a) Schematic plots of the hinge-state network (dark green lines) in prismatic geometry with even QSHI layers. The hinge states on the top and bottom surfaces reside on the same side of the glide plane (light green plane).
%(b) The hinge-state network with odd QSHI layers by removing the topmost layer B. The hinge states on the top surface shift to the other side.
%In either case, the helical hinge states propagate around the whole sample.
%} \label{fig-hinge}
%\end{figure}

\section{Minimal model}
\begin{figure}[t!]
\centering
\includegraphics[width=1\linewidth]{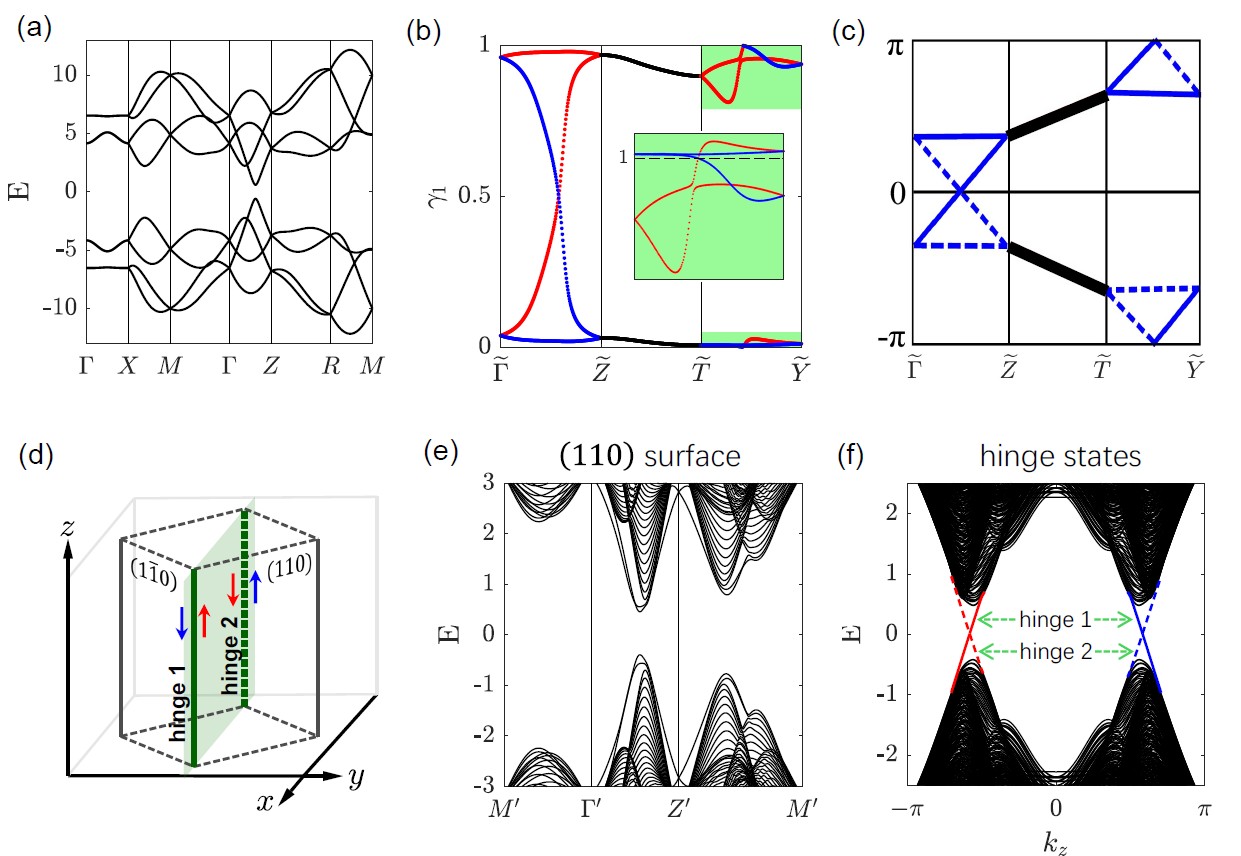}
\caption{(Color online) 
(a) Bulk energy bands of the gTCI model in Eq.~\pref{Hamiltonian}. The parameters are $-m=t=2$, $v_x=v_y=A_x=A_y=2$, $t_1=t_4=0.5$, $t_5=t_2=2$ and $t_6=-t_3=1.5$.
%(b) WCCs $\gamma_1$ of the $k_z$-directed Wilson loop of the four-occupied bands with fixed $k_y=0$. The WCCs form two separated parts of Wilson bands localized around $\gamma_1=0$ and $\gamma_1=0.5$, related by $g_y$.
(b) WCCs $\gamma_1$ of the $k_x$-directed Wilson loops for the four occupied bands. Red and blue dots carry different glide eigenvalues $+ie^{-i k_z/2}$ and $-ie^{-i k_z/2}$, respectively. %The zigzag pattern of hourglass Wilson bands characteristic of the gTCI phase is observed. 
Inset plot shows the zoom-in of the WCCs along $\widetilde{T}$-$\widetilde{Y}$ highlighted in green, showing the hourglass-type pattern.
%(c) Nested WCCs $\gamma_{2}$ of the $k_y$-directed nested Wilson loop for the Wilson bands around $\gamma_1=0$ (inside the blue frame). The nontrivial winding pattern shows a $\mathbb{Z}_2=1$ QSHI on $(001;0)$ layer.
(c) Schematic plot: The zigzag-connected hourglass-type Wilson bands for a gTCI, adapted from Ref.~\cite{hourglass-fermion}. Solid and dashed lines carry different glide eigenvalues.
(d) Schematic plot of a rod geometry (along the $z$ direction), bounded by the (110) and (1$\bar{1}$0) surfaces.
(e) Surface spactra of the (110) and (1$\bar{1}$0) surfaces, which are identical due to $g_y$.
(f) $k_z$-directed spectra of the rod geometry in panel (d). The topological helical states are located at the hinges.
%(g) Schematic plots of the helical hinge states (dark green lines) for a gTCI in a prismatic geometry, with an even number of QSHI layers. The hinge states on the top and bottom surfaces lie on the same side of the glide plane (light-green plane).
%(h) Hinge-state configuration with an odd number of QSHI layers, obtained by removing the topmost QSHI layer. The hinge states on the top surface shift to the opposite side.
%In both cases, the helical hinge states propagate around the entire sample.
} \label{fig-model}
\end{figure}

%\paragraph*{Minimal model.}
To study the $T$-DAXI in nonsymmorphic TCI, we take gTCI as an example and construct an eight-band model,
\begin{equation}\label{Hamiltonian}\begin{aligned}
    &H_g(\bk) = \left(\begin{array}{cc} H_+(k_x,k_y) & H_{1}(k_z) \\ H_{1}^\dagger(k_z) & H_-(k_x,k_y) \end{array}\right),\\
    &H_\pm(k_x,k_y)=M_ks_0\sigma_z+v_x\sin{k_x}s_z\sigma_x-v_y\sin{k_y}s_0\sigma_y\\
   &\qquad\qquad\qquad \pm A_x\sin{k_x}s_y\sigma_x \pm A_y\sin{k_y}s_x\sigma_x, \\
   &M_k=m+t\cos{k_x}+t\cos{k_y}, \\
    &H_{1}(k_z)=e^{i k_z/2}\left(f_1\cos{\frac{k_z}{2}}+f_2\sin{\frac{k_z}{2}}\right).
\end{aligned}\end{equation}
Here, $\tau$, $s$ and $\sigma$ are Pauli matrices for layer, spin and orbital degrees of freedom, respectively. $H_{+}$ and $H_-$ describe the two QSHI layers (with $|m|<|2t|$) in a unit cell, respectively located at $(001;d_0)$ and $(001;d_0+c/2)$, and related by the glide symmetry $g_y$. We choose $d_0=0$ and $c=1$ without loss of generality. $H_{1}(k_z)$ represents the interlayer coupling with $f_1=t_1s_0\sigma_0+it_2s_y\sigma_x+it_3s_x\sigma_0$ and $f_2=it_4s_0\sigma_x+t_5s_y\sigma_0+t_6s_x\sigma_x$.
The energy bands of the gTCI model $H_g(\bk)$ are shown in Fig.~\ref{fig-model}(a).

%\section{Nested Wilson loop method}
%\paragraph*{Nested Wilson loop method.}
If $H_{1}(k_z)$ is omitted, based on the real-space layer construction, the model realizes a gTCI with invariants $\delta_h=1$ and $(\nu_0;\nu_1\nu_2\nu_3)=(0;000)$~\cite{LC-Song}. Upon adding $H_{1}(k_z)$ without closing the bulk gap, these invariants remain unchanged. In addition, the gTCI can also be confirmed by the zigzag pattern of hourglass Wilson bands on a glide-preserving surface. To illustrate this, we compute the glide-protected $k_x$-directed Wilson loop and plot the Wannier charge centers (WCCs) along the (100)-surface Brillouin-zone path $\widetilde{\Gamma}$$\widetilde{Z}$$\widetilde{T}$$\widetilde{Y}$ in Fig.~\ref{fig-model}(b). One can see that the Wilson bands have the nontrivial zigzag pattern, which is identical to Fig.~\ref{fig-model}(c) adapted from Ref.~\cite{hourglass-fermion}, confirming a gTCI phase.
Meanwhile, the glide-protected Wilson loop carries a glide $Z_4$ value~\cite{Aris-PRX,Z4value-chiy,chiy-value-PRB} that can be computed as
\begin{equation}\label{chi_y}\begin{aligned}
    \chi_y=&\frac{1}{2\pi}\sum_{j=1}^{n_{\mathrm{occ}}/2}\left[ 2\gamma^-_{1,j}(\widetilde{Z}) - 2\gamma^-_{1,j}(\widetilde{T}) + \int_{\widetilde{\Gamma}}^{\widetilde{Z}}(d\gamma^+_{1,j}-d\gamma^-_{1,j}) \right. \\
    & \left. + \int_{\widetilde{T}}^{\widetilde{Y}}(d\gamma^+_{1,j}-d\gamma^-_{1,j}) \right] + \frac{1}{2\pi}\sum_{j=1}^{n_{\mathrm{occ}}}\int_{\widetilde{Z}}^{\widetilde{T}}d\gamma_{1,j}\ \mathrm{mod}\ 4,
\end{aligned}\end{equation}
with $\gamma^{\pm}_1$ being the WCCs of the $k_x$-directed Wilson loops in the $\pm$ glide sectors, respectively. The glide $Z_4$ invariant is obtained to be $\chi_y=2$, further confirming the gTCI phase. %Therefore, the spin-resolved topology in the $s_y$ direction yields a $T$-DAXI as discussed above.

%On the other hand, the hourglass invariant of the model can be computed by the nested Wilson loop method. For the four occupied bands, the eigenvalues $e^{i2\pi\gamma_1(k_x,k_y)}$ of $k_z$-directed Wilson loops ($W_1(k_x,k_y)$) determine the Wannier charge centers (WCCs) in the $z$ direction. Fig.~\ref{fig-model}(b) shows the WCCs $\gamma_1(k_x,k_y)$, which form two separated parts of Wilson bands. One part of Wilson bands is located near $z=0$ and the other part is near $z=0.5$. Then, one can apply $k_y$-directed nested Wilson loops for each part, whose eigenvalues are defined as $e^{i2\pi\gamma_{2}(k_x)}$. In Fig.~\ref{fig-model}(c), for the two Wilson bands around $z=0$, the nested Wilson loop shows a nontrivial pattern with $\mathbb{Z}_2=1$, indicating a QSHI on $(001;0)$ layer. Considering the $g_y$ symmetry, we conclude that the model has $\delta_h=1$, belonging to a gTCI.

On glide-preserved surfaces, the hourglass-like surface states are expected in a gTCI. However, on the (110) surface without glide symmetry, a glide-breaking perturbation introduces a mass term that gaps the surface spectrum. 
Meanwhile, due to the glide symmetry, the (1$\bar{1}$0) and (110) surfaces have the same spectrum, but with opposite mass terms. Consequently, helical hinge states emerge at the interface between the two surfaces, as shown in Fig.~\ref{fig-model}(f). The presence of the topological helical hinge states demonstrates that a glide TCI also manifests itself as a helical HOTI.
We further find that the helical hinge states extend to propagate through the top and bottom surfaces, encircling the entire sample of a glide TCI. The position of the top and bottom helical hinge states is related to the even or odd number of total QSHI layers of the sample, as shown in the insets of Fig.~\ref{fig-axion}(c) and (d).
%analogous to the chiral hinge states in antiferromagnetic MnBi$_2$Te$_4$.
We note that the helical hinge modes that encircle the whole sample of a glide TCI require nearly perfect glide symmetry of the real sample.%, which is delicate for a real sample of a material. For example, surface reconstruction or disorder may close the side-surface band gap of a real sample (equivalent to “gluing” QSHI layers to the surface), and remove some of   the hinge modes.

%An effective boundary Hamiltonian analysis further shows that a nonsymmorphic TCI itself realizes a helical HOTI, hosting a pair of helical hinge states encircling the entire sample.

\section{Properties of a $T$-DAXI}
The spin-resolved topology of our gTCI model in the $s_y$ direction thus yields a $T$-DAXI as discussed above. In this section, we study the properties of the $T$-DAXI in three aspects using the gTCI model, including partial axion angles, surface partial quantum anomalies, and spin-magnetoelectric effect.

\subsection{Partial axion angles}
%\paragraph*{Partial axion angles.---}
As viewed as two $T$-related copies of a magnetic AXI, the $T$-DAXI can be characterized by partial axion angles $\theta^\pm$, which can be evaluated from partial Wannier functions constructed in separated spin subspaces~\cite{axionangle-WCC-2020}. In this scenario, for an $L$-unit-cell slab along the $z$ direction, we define an auxiliary slab partial axion term,
%Directly calculating partial axion angles for the $T$-DAXI in an inversion TCI is challenging due to the coexistence of a partial Chern insulator~\cite{the-nc}. However, such a hardship is ruled out here in nonsymmorphic TCIs. 
\begin{equation}\label{partial-axion-angle}\begin{aligned}
    \theta^{\pm}_{slabL}=-\frac{1}{L}\int{d^2\bk}\sum_{n=1}^{L\cross N_{\mathrm{occ}}/2}\left[ z^{\pm}_{n\bk}\tilde{\Omega}_{\bk n}^{xy,\pm} \right],
\end{aligned}\end{equation}
where $z^\pm_{n\bk}$ are the eigenvalues of $z$-directed position operator defined in the spin subspaces $P^{\pm}_{\bk}\hat{z}P^{\pm}_{\bk}$, where $P^{\pm}_{\bk}=\sum_{n}\ket{\psi^{\pm}_{n\bk}}\bra{\psi^{\pm}_{n\bk}}$ project onto the two spin subspaces.
$\tilde{\Omega}_{\bk n}^{xy,\pm}$ is the corresponding partial non-Abelian Berry curvature~\cite{axionterm-prl,zoperator-1,zoperator-2,non-Abelian-Berry} defined in the eigenstates of $P^{\pm}_{\bk}\hat{z}P^{\pm}_{\bk}$.
$N_{\mathrm{occ}}$ is the number of occupied bands. In the infinite thickness limit, the slab partial axion terms in Eq.~\pref{partial-axion-angle} converge to the partial axion angles defined by the Chern-Simons three-form~\cite{axionangle-WCC-2020} in the two spin subspaces, \ie $\theta^{\pm}=\lim_{L\rightarrow \infty}\theta^{\pm}_{slabL}$.  

\begin{figure}[tb!]
\centering
\includegraphics[width=1\linewidth]{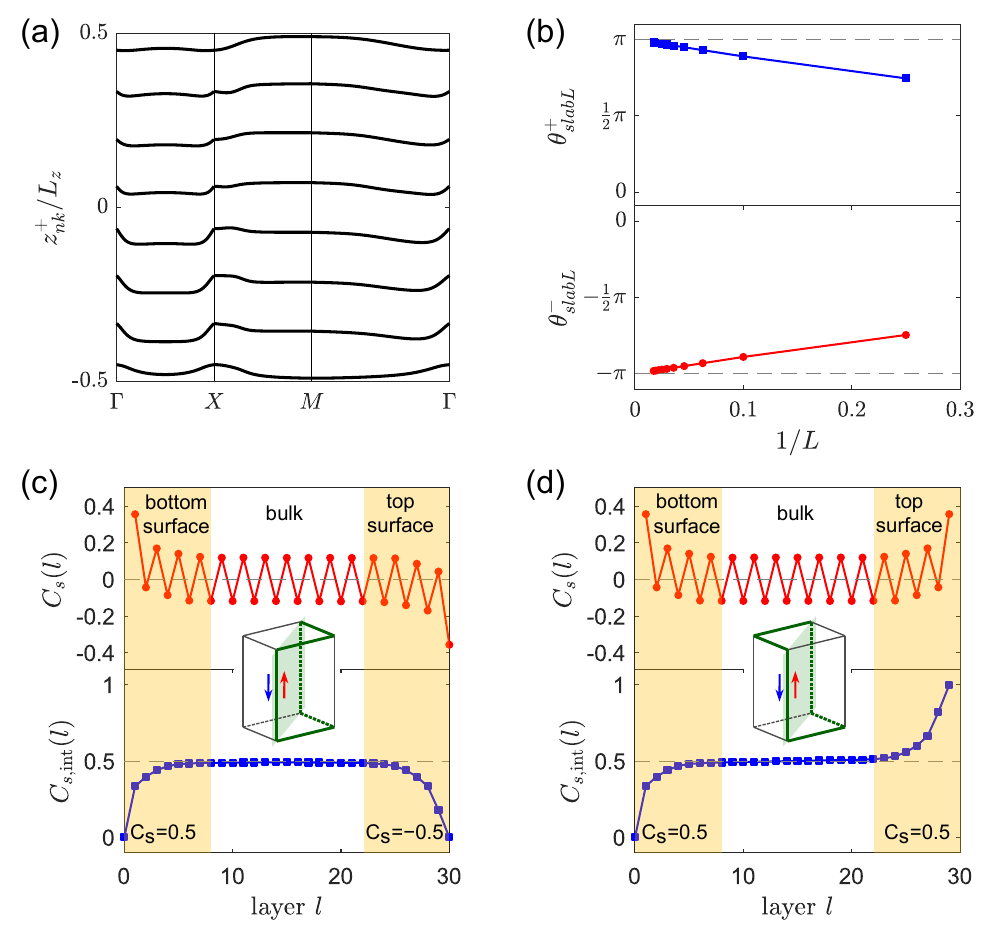}
\caption{(Color online) 
(a) Partial WCCs $z^{+}_{n\bk}$ for the $s_y$ upper spin bands of a 4-unit slab, where 8 upper spin Wannier bands are located on 8 layers (two QSHI layers per unit cell).
(b) Evolution of the slab partial axion terms $\theta^\pm_{slabL}$ in the $s_y$ resolution as a function of $1/L$. $\theta^{\pm}_{slabL}$ are quantized to $\pm\pi$ as $L\rightarrow\infty$, indicating a $T$-DAXI phase.
(c) Upper panel: layer-dependent spin Chern number $C_{s}(l)$ for a 15-unit-cell slab (being even QSHI layers). Lower panel: accumulated spin Chern number $C_{s,\mathrm{int}}(l)$ up to depth $l$.
(d) Layer-dependent spin Chern number and its accumulation for a modified slab after removing the topmost QSHI layer (being odd QSHI layers).
Insets of (c) and (d): Helical hinge states configurations (green lines) for even- and odd-layer-QSHI geometry, respectively, which serve as conducting channels for the surface SHC.
} \label{fig-axion}
\end{figure}

For illustration, in the $s_y$ resolution, we show $z^{+}_{n\bk}$ for the upper spin bands of a 4-unit-cell slab of our gTCI model in Fig.~\ref{fig-axion}(a). Eight separated partial WCCs correspond to the eight weakly coupled QSHI layers. In Fig.~\ref{fig-axion}(b), the upper (lower) panel shows the evolution of $\theta^{+}_{slabL}$ ($\theta^{-}_{slabL}$) as a function of the inverse of $L$. The intercepts on the $y$ axis indicate $\theta^{\pm}=\pm \pi$, confirming the $T$-DAXI phase. Here, the signs ``$\pm$'' before $\pi$ are still well defined by means of the asymptotic convergence behavior together with time-reversal symmetry.

\begin{figure}[t!]
\centering
\includegraphics[width=1\linewidth]{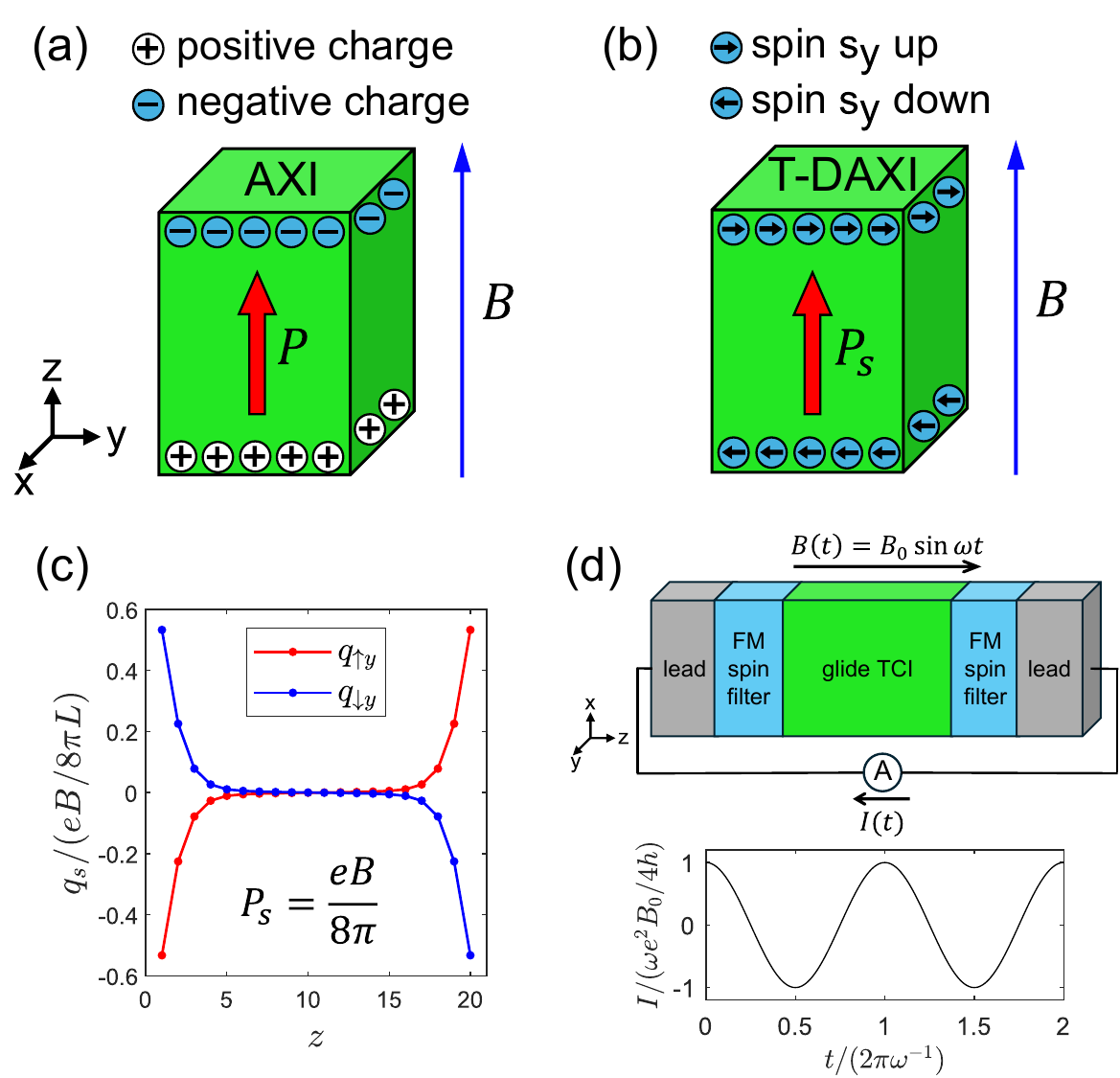}
\caption{(Color online) 
(a) Topological charge polarization induced by an external magnetic field in a magnetic AXI.
(b) Topological nonlocal spin polarization induced by an external magnetic field in a $T$-DAXI.
(c) For $s_y$-conserved parameters by setting $v_x=A_y=t_3=t_6=0$ and $t_1=t_4=1.5$, the $s_y$ spin distributions in each unit cell are shown along the $z$ direction under a $z$-directed magnetic field. Red and blue dotted lines indicate $s_y$ up and $s_y$ down, respectively. The total nonlocal spin polarization reaches the quantized value $P_{s}=eB/8\pi$.
(d) Alternating spin current in a gTCI tunneling junction induced by a time-periodic external magnetic field.
Using $y$-directed polarized ferromagnetic (FM) spin filters, the spin-valve geometry can filter and convert the spin current into measurable charge current.
} \label{fig-spin polarization}
\end{figure}

\subsection{Layer-dependent spin Chern numbers}
%\paragraph*{Layer-dependent spin Chern numbers.---}
We now discuss the surface anomaly implied by the partial axion angles. In a magnetic AXI, the surface anomalous Hall conductivity (AHC) is half-quantized as  $\sigma^{\mathrm{AHC}}=e^2/2h$~\cite{Wilczek-AXI,Essin-AXI,AXI-Vanderbilt-2018,Olsen-AXI}. Regarding a $T$-DAXI, the net surface AHC vanishes because of the time-reversal symmetry. However, with spin conservation, the topological contribution to the surface SHC should be nontrivial and half-quantized as $\sigma^{\mathrm{SHC}}=e/4\pi$~\cite{the-nc}. 
This half-quantized surface SHC feature can be captured using the layer-dependent spin Chern number $C_s(l)=\frac{1}{2}\left[ C^{+}(l)-C^{-}(l) \right]$, where $C^\pm(l)$ resolve the Thouless-Kohmoto-Nightingale-Nijs integer into contributions from individual layers~\cite{Essin-AXI},
\begin{equation}\begin{aligned}
    C^{\pm}(l)=\frac{i}{2\pi}\int d^2k\text{Tr}\left[ P^{\pm}_\bk\epsilon_{ij} \left(\partial_i P^{\pm}_\bk\right) P_{l} \left(\partial_j P^{\pm}_\bk\right) \right].
\end{aligned}\end{equation}
Here, $\epsilon_{ij}$ is the antisymmetric tensor with $i,j\in\{k_x,k_y\}$. $P_{l}$ projects onto the $l$-th layer of a slab ($l\in[1,2L]$). 
In the upper panel of Fig.~\ref{fig-axion}(c), we plot the layer-dependent spin Chern number $C_s(l)$ in the $s_y$ resolution for an $L$-unit-cell slab of the gTCI model. $C_s(l)$ is primarily localized on the surfaces and decays to zero in the bulk in an oscillatory manner. We then accumulate the layer-dependent spin Chern number to depth $l$ according to $C_{s,\mathrm{int}}(l)=\sum_{l'=1}^{l}\bar{C}_{s}(l')$, with $\bar{C}_s(l')=\frac{1}{2}\left[ C_s(l')+C_s(l'+1) \right]$ being the average of two neighboring layers~\cite{AXI-Vanderbilt-2018}. The results are plotted in the lower panel of Fig.~\ref{fig-axion}(c). Although the total spin Chern number of the slab vanishes, the spin Chern numbers on the bottom and top surfaces are half-quantized to $C_{s,\text{bot}}=-C_{s,\text{top}}=1/2$, consistent with the surface partial quantum anomaly of a $T$-DAXI.
Moreover, if we modify the slab by removing the topmost QSHI layer (a half unit cell) from the slab (resulting in $l\in [1,2L-1]$), the half-quantized plateau on the remaining surfaces persists while the sign of $C_{s,\text{top}}$ flips (\ie $C_{s,\mathrm{top}}=1/2$). And consequently, the total spin Chern number of the modified slab becomes nonzero, $C_s=1$.

\begin{figure*}[tb!]
\centering
\includegraphics[width=1\linewidth]{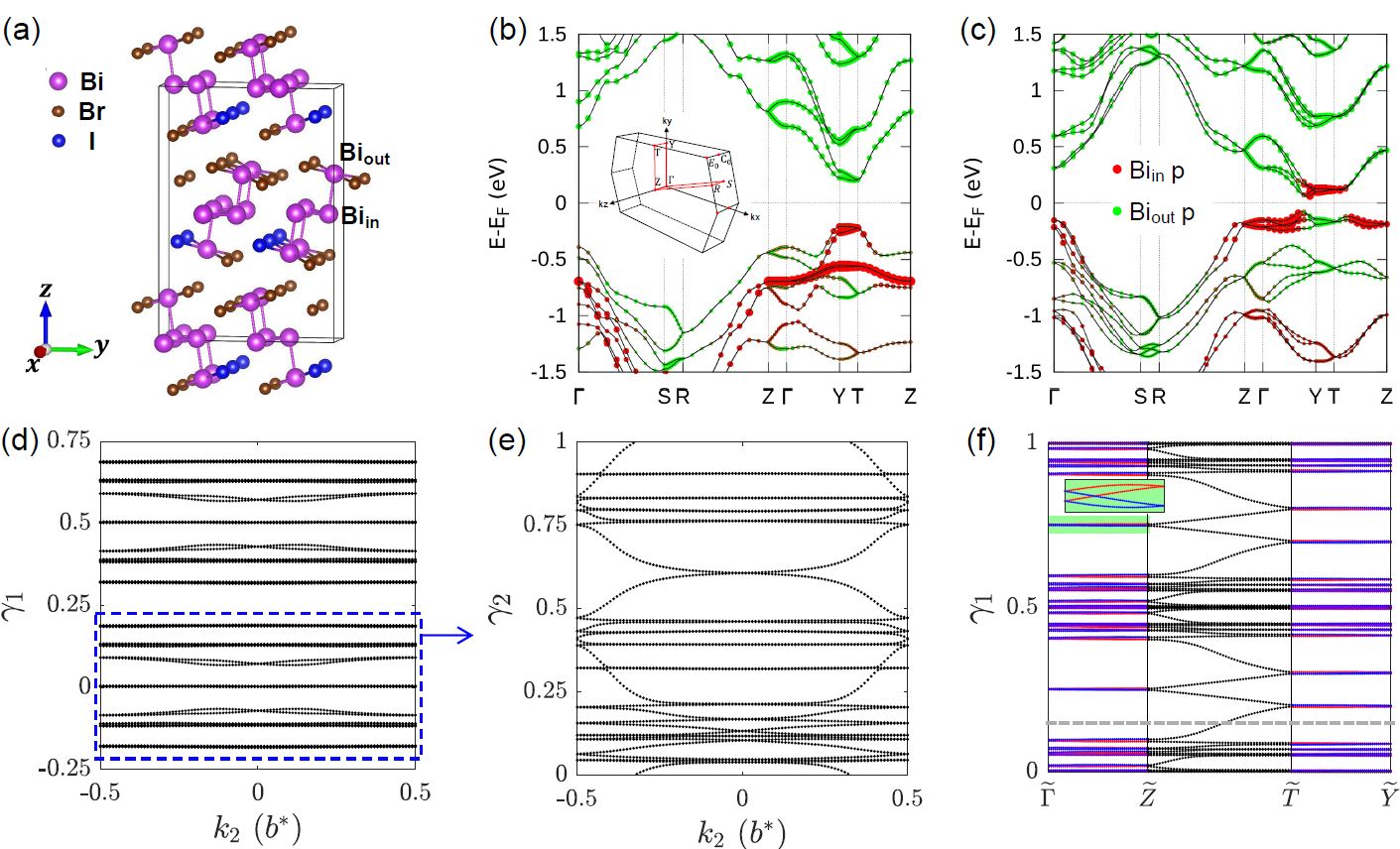}
\caption{(Color online) 
(a) Crystal structure of Bi$_4$Br$_3$I.
(b) Electronic band structure without SOC. Inset: Brillouin zone and high-symmetry paths.
(c) Electronic band structure with SOC. The size of the red (green) circles indicate the $p$-orbital components of two kinds of Bi atoms, Bi$_\mathrm{in}$ and Bi$_\mathrm{out}$ labeled in (a). 
(d) WCCs of the $k_z$-directed Wilson loops of the occupied bands.
(e) The nested Wilson loops for the $k_z$-directed Wilson bands around $\gamma_1=0$, depicted by a dashed-blue frame in (d). It shows a zigzag pattern indicating the QSHI phase.
(f) $k_x$-directed Wilson loops for the occupied bands, carrying the glide $Z_4$ value $\chi_y=2$. Red and blue dots carry different glide eigenvalues $+ie^{-i k_z/2}$ ($+$ glide sector) and $-ie^{-i k_z/2}$ ($-$ glide sector) on both $\widetilde{\Gamma}\widetilde{Z}$ and $\widetilde{T}\widetilde{Y}$ paths, respectively. The inset shows the zoom in highlighted in green.
With a horizontal reference line (dashed-gray line), the $\chi_y$ in Eq.~\pref{chi_y} can be computed by simply counting the number of times of the Wilson bands crossing the line, \ie $\chi_y=S^+ - S^- + 2S_{\widetilde{Z}\widetilde{T}}$. $S^\pm$ is defined as the number of positively sloped crossings minus the number of negatively sloped crossings in the \(\pm\) glide sectors along both $\widetilde{\Gamma}\widetilde{Z}$ and $\widetilde{T}\widetilde{Y}$. $S_{\widetilde{Z}\widetilde{T}}$ is the number of crossings on $\widetilde{Z}\widetilde{T}$, with a factor of 2 due to the double degeneracy.
%The WCCs around $\gamma_1=0.5$ should show the same nested Wilson loop pattern due to $g_y$ or $s_2$. These form a QSHI layer construction on the $(001;0)$ and $(001;\frac{\bc}{2})$ planes, indicating a nonsymmorphic TCI with $\delta_h=\delta_s=1$ and $(\nu_0;\nu_1\nu_2\nu_3)=(0;000)$.
%(f) Spin bands formed by eigenvalues of $\bS_y(\bk)$, where a spin-band gap is clearly observed.
} \label{fig-BiBr}
\end{figure*}

\subsection{Spin-magnetoelectric effect}
%\paragraph*{Spin-magnetoelectric effect.---}
For a magnetic AXI, the quantized axion angle $\theta=\pi$ implies a topological magnetoelectric effect: an applied magnetic field $B$ induces a quantized charge polarization $\bP=e^2\bB/2 h$ parallel to the magnetic field, so that electrons accumulate on one surface from the opposite surface~\cite{AXI-Qi-prb,dynamical-AXI-Qi-prb} [see Fig.~\ref{fig-spin polarization}(a)]. In a $T$-DAXI, the partial axion angles $\theta^\pm=\pm\pi$ lead to a spin-magnetoelectric effect. Under a magnetic field, the net charge response vanishes because of time-reversal symmetry, but electrons carrying opposite spins will accumulate on two opposite surfaces, producing a nonlocal spin polarization parallel to the external magnetic field, as shown in Fig.~\ref{fig-spin polarization}(b).

To calculate the real-space nonlocal spin polarization in the gTCI model with parameters modified to $s_y$ conservation, we consider an $L$-unit-cell slab under a static magnetic field $\bB=(0,0,B)$ applied along the $z$ direction, which contributes the Peierls substitution $e^{(ie/\hbar)\int d\bl\cdot \bA}$ with $\bA=(-yB,0,0)$. We obtain the $s_y$ spin distribution via $q_{s}(z)=\frac{\hbar}{2V}\langle \hat{N}_{s}(z) \rangle$, where $\hat{N}_{s}(z)$ with $s=\uparrow_y$ and $\downarrow_y$ are the total-density operators at position $z$ for spin up and spin down of $s_y$, respectively. $V$ is the volume of the slab. In Fig.~\ref{fig-spin polarization}(c), we show the $s_y$ spin distributions along the $z$ direction $q_s(z)$ for a 20-unit-cell slab, measured relative to an average $\ovl{q}_{s}=\sum_zq_{s}(z)/L$. One clearly sees that $s_y$-up electrons transfer from the bottom surface to the top surface while $s_y$-down electrons transfer in the opposite direction, producing a real-space nonlocal spin polarization. We further quantify the nonlocal spin polarization by $P_s=\sum_z z[q_{\uparrow_y}(z)-q_{\downarrow_y}(z)]$ and it quantizes to $P_s=eB/8\pi$, determined by the partial axion angles. Moreover, if the applied magnetic field is time-periodic with a frequency $\omega$, $B(t)=B_0\sin{\omega t}$, the nonlocal spin polarization should also be periodic and thus generate an alternating spin current with $I_s=dP_s(t)/dt=\omega eB_0\cos(\omega t)/8\pi$ in a tunneling junction.
Using $y$-directed polarized ferromagnetic (FM) spin filters, the spin-valve geometry can filter and convert the spin current into measurable charge current, as shown in Fig.~\ref{fig-spin polarization}(d).
This provides a direct transport signature of the spin-magnetoelectric effect in a $T$-DAXI.
When spin conservation is relaxed, although the $T$-DAXI phase is still well-defined without closing the spin gap, the spin-magnetoelectric response should deviate from but lie close to the quantized value if the spin-nonconserving SOC is relatively weak (see details in Appendix~\ref{SM: spin-magnetoelectric-effect}).

\begin{figure*}[!t]
\centering
\includegraphics[width=1\linewidth]{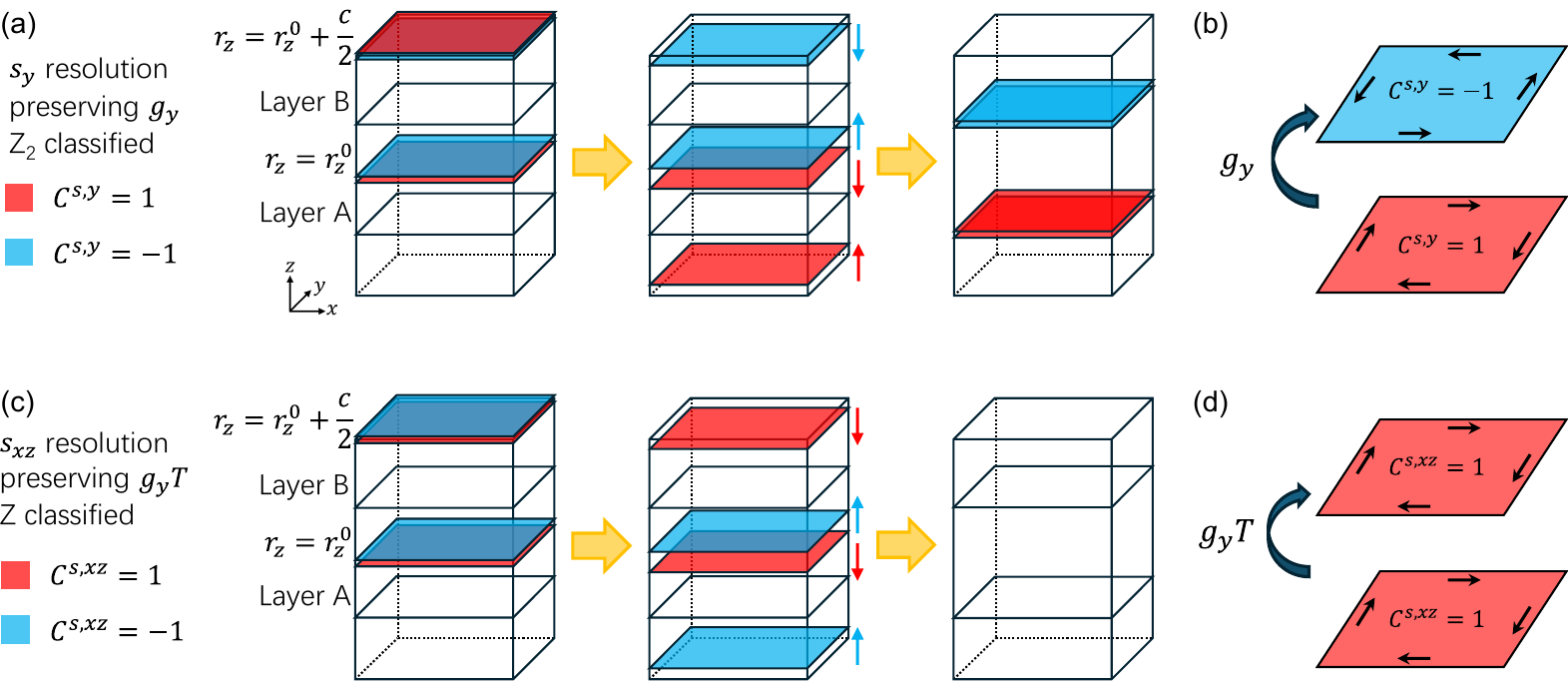}
\caption{(Color online) 
In a $T$-invariant 3D insulator with glide $g_y$, bubble-equivalence processes for (a) the $g_y$-preserved $s_y$-projected spin subspaces and (c) the $g_yT$-preserved $s_{xz}$-projected spin subspaces. (b,~d) Relations of the spin Chern number under symmetries $g_y$ and $g_yT$.
} \label{fig-bubble}
\end{figure*}

%\section{Application to $\gamma$-B\lowercase{i}$_4$B\lowercase{r}$_4$}
\section{Application to B\lowercase{i}$_4$B\lowercase{r}$_{x}$I$_{4-x}$}\label{Application to material}
%\paragraph*{Application to $\gamma$-Bi$_4$Br$_4$.---}
The mixed bismuth monohalides Bi$_4$Br$_{x}$I$_{4-x}$~($x=1,3$)  have been synthesized experimentally~\cite{BiBrI-exp}, which crystallize in a base-centered structure belonging to the space group $Cmc2_1$ (\#SG 36). We take Bi$_4$Br$_3$I of Fig.~\ref{fig-BiBr}(a) as an example in the main text.
The chains extend along the $a$ axis and assemble into layers within the $ab$ plane. 
Two adjacent van der Waals (vdW) layers form an AB stacking arrangement and are related by both glide $g_y$ and screw $s_2$ symmetries, while an interlayer sliding breaks the inversion symmetry of the bulk crystal.
The electronic band structure of Bi$_4$Br$_3$I without and with SOC is plotted in Fig.~\ref{fig-BiBr}(b) and (c), respectively, both showing a global bulk gap. Upon including SOC, band inversion occurs along the $Y$-$T$ path in the Brillouin zone. The orbital-projected band structures reveal that this band inversion involves bands with Bi $p$-orbital character at different positions [Bi$_\mathrm{in}$ and Bi$_\mathrm{out}$ in Fig.~\ref{fig-BiBr}(a)].
Due to the time-reversal symmetry, Bi$_4$Br$_3$I can be classified by four $\mathbb{Z}_2$ topological invariants. A Wilson loop calculation yields $(\nu_0;\nu_1\nu_2\nu_3)=(0;000)$, indicating that the compound is neither a strong nor a weak 3D TI.
Nevertheless, in Fig.~\ref{fig-BiBr}(d), we show WCCs $\gamma_1$ of the $k_z$-directed Wilson loops of the occupied energy bands. Two sets of isolated WCCs are located around $\gamma_1=0$ and $\gamma_1=0.5$. Fig.~\ref{fig-BiBr}(e) further shows the $k_1$-directed nested Wilson loop for the WCCs around $\gamma_1=0$ [inside the dashed-blue frame of Fig.~\ref{fig-BiBr}(d)]. A nontrivial zigzag pattern indicates a $\mathbb{Z}_2=1$ QSHI layer at $(001;0)$ plane. Due to $g_y$ and $s_2$, another QSHI layer is located at $(001;c/2)$. By real-space layer construction, this reveals a nonsymmorphic TCI protected by both $g_y$ and $s_2$, as illustrated in Fig.~\ref{fig-LC}(a,c). In short, our calculations show that Bi$_4$Br$_3$I is a nonsymmorphic TCI with $\delta_h=\delta_s=1$ and $(\nu_0;\nu_1\nu_2\nu_3)=(0;000)$.
In addition, we calculate the glide-protected $k_x$-directed Wilson loop in Fig.~\ref{fig-BiBr}(f) carrying the glide $Z_4$ value $\chi_y=2$, which indicates a gTCI phase.
Furthermore, we illustrate the $s_y$ spin spectrum along high-symmetry path in Fig.~\ref{fig-sNWL}(a), and calculate the global minimum spin gap in the whole Brillouin zone. The minimum spin gap is obtained to be $\Delta_{s_y}=0.33$, which is larger than that ($\Delta_{s_x}=0.26$) of $\alpha$-Bi$_4$Br$_4$~\cite{the-nc}.
Therefore, in the $s_y$ spin direction, Bi$_4$Br$_3$I realizes a $T$-DAXI phase. To further verify the spin-resolved topology, we have performed the  spin-resolved nested Wilson loop calculations (see Appendix \ref{SM: sNWL}), whose results agree with the spin-resolved topological classification in nonsymmorphic TCIs discussed in Section \ref{section: spin-resolved topology}.
Moreover, a new bulk phase of Bi$_4$Br$_4$ and Bi$_4$I$_4$---denoted as the $\gamma$ phase---is theoretically predicted to be dynamically stable~\cite{BiBr-shh} and also realizes $T$-DAXI in our calculations.
%The Total energy of $\gamma$-Bi$_4$Br$_4$ is lower than the experimentally synthesized $\alpha$ phase~\cite{BiX-zaac.19784380105,BiX-FILATOVA20071103,BiBr-1layer.nl,BiX-NM2021}. Therefore, the synthesis of this new $\gamma$ phase is accessible.Since $\gamma$-Bi$_4$Br$_4$ and $\gamma$-Bi$_4$I$_4$ also belong to the space group $Cmc2_1$, we carry out similar Wilson loop calculations and show that they are nonsymmorphic TCIs exhibiting $T$-DAXIs. 

\section{Conclusion and discussion}
%\paragraph*{Conclusion and discussion.---}
In summary, we have identified spin-resolved topological classifications and revealed $T$-DAXIs in nonsymmorphic TCIs. In certain spin-resolved directions, the $T$-DAXI is protected by nonsymmorphic symmetries. 
Explicitly, the $g_y$- and $s_2T$-protected $T$-DAXI are indicated by the hourglass invariant $\delta_{h}=1$ and the screw invariant $\delta_{s}=1$, respectively, with trivial time-reversal $\mathbb{Z}_2$ indices $(\nu_0;\nu_1\nu_2\nu_3)=(0;000)$.
%These nonsymmorphic TCIs exhibit helical hinge states that surround the whole sample, resembling higher-order topological insulators. 
The nonsymmorphic $T$-DAXI is revealed to be characterized by several spin-resolved features: quantized partial axion angles, half-quantized surface spin Chern numbers, and topological nonlocal spin polarizations. Some novel experimentally accessible signatures of the $T$-DAXI can then be proposed.
The glide/screw-protected helical hinge states encircling the entire sample provide a spectroscopic signature accessible to angle-resolved photoemission spectroscopy or scanning tunneling microscopy.
The surface quantum anomaly produces opposite, topologically contributed nonzero surface spin Hall conductivities, which are half-quantized to $-\sigma^{s,y}_{\text{top}}=\sigma^{s,y}_{\text{bot}}=e/4\pi$ on the top and bottom surfaces with spin conservation. This can be probed in a spin-transport experiment.
Due to the spin-magnetoelectric effect, a time-periodic external magnetic field can generate an alternating spin current in a tunnel junction, providing a dynamical transport probe of the partial axion physics.
These experiments can be performed in the mixed bismuth monohalides Bi$_4$Br$_{x}$I$_{4-x}$ and in other TCIs with nonsymmorphic symmetries.
Furthermore, by changing the slab sample between even and odd layer counts, the total spin Chern number of a gTCI/sTCI can be controlled to vanish or quantize to 1, and the response of the system accordingly alternates between a $T$-DAXI and a QSHI. This layer sensitivity suggests a route towards layer-manipulated ``layertronics''.
%In addition, changing a slab of a nonsymmorphic TCI from even to odd layers swaps the hinge-state locations and flips the sign of the surface topologically-contributed half-quantized SHC (and the associated nonlocal spin polarization). This suggests straightforward layer control of $T$-DAXI responses --- a promising route toward layer-manipulated ``layertronics''. Moreover, many experimentally observed responses in magnetic AXIs (for example, antiferromagnetic MnBi$_2$Te$_4$) --- and the corresponding probes (transport, magnetoelectric and optical signatures) --- should be revisited and generalized for $T$-DAXIs in nonsymmorphic TCIs. 
In addition, although this work focuses on $T$-invariant systems, the symmetry arguments should naturally generalize to magnetic gTCIs and sTCIs. In those magnetic nonsymmorphic TCIs, one can expect that chiral hinge states appear surrounding the whole samples, and that the axion angles are quantized to $\pi$ by the nonsymmorphic symmetries, with quantized magnetoelectric responses measurable without involving gapless boundary states.
Finally, we anticipate that this work will motivate further experimental and theoretical studies on material realizations and device concepts that exploit partial axion physics.

\ \\
\section*{Acknowledgments}
This work was supported by the National Natural Science Foundation of China (Grant No. 12188101), the National Key R\&D Program of China (Grant No. 2022YFA1403800),  and the Center for Materials Genome. And H. W. acknowledges support from the New Cornerstone Science Foundation through the XPLORER PRIZE.

\appendix

\begin{figure*}[!t]
\centering
\includegraphics[width=1\linewidth]{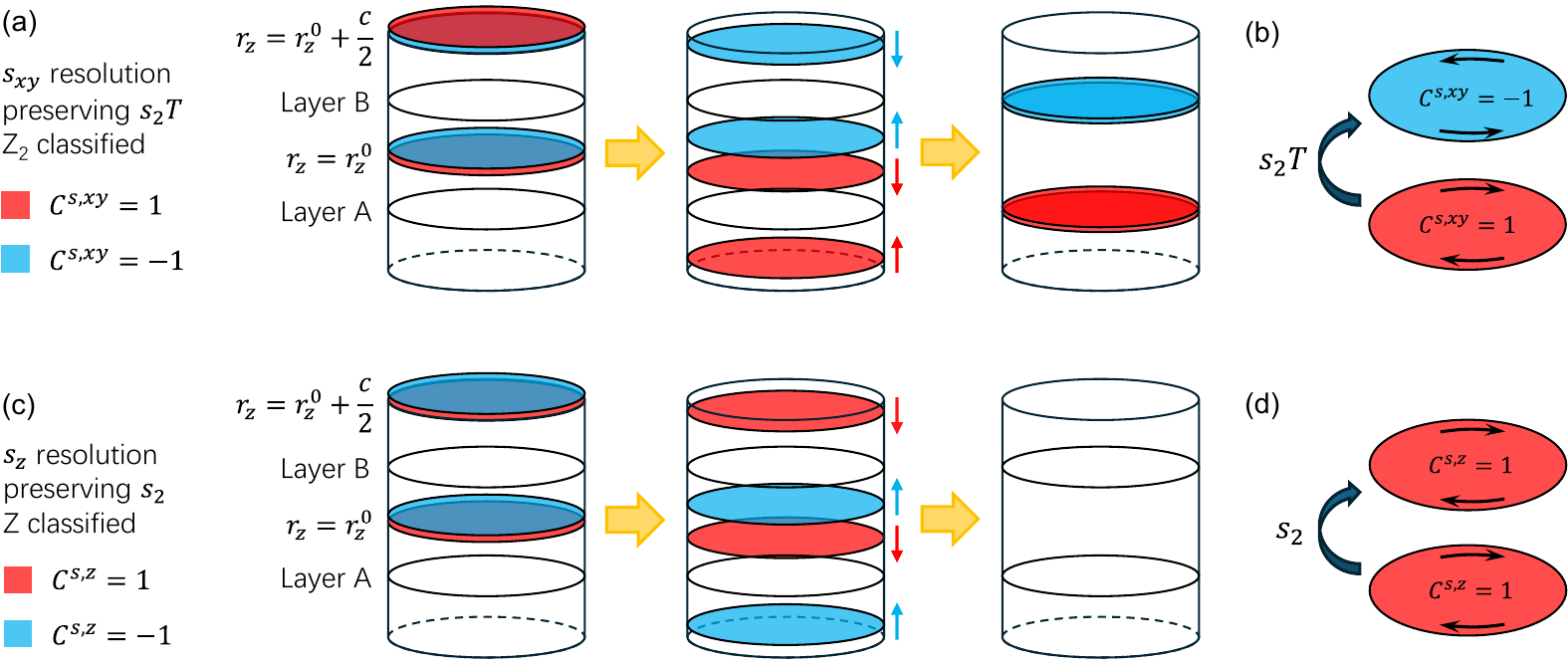}
\caption{(Color online) 
In a $T$-invariant 3D insulator with screw $s_2$, bubble-equivalence processes for (a) the $s_2T$-preserved $s_{xy}$-projected spin subspaces and (c) the $s_2$-preserved $s_{z}$-projected spin subspaces. (b,~d) Relations of the spin Chern number under symmetries $s_2T$ and $s_2$.
} \label{fig-bubble-screw}
\end{figure*}

\section{Bubble equivalence and spin-resolved topological classification}
\label{SM: bubble-equvalence}
We employ the real-space topological classification method, bubble-equivalence analysis~\cite{bubble-equivalence-2020}, to classify the spin-resolved layer constructions. In this approach, a bubbling process refers to a symmetry-preserving deformation in which topologically trivial layers are added to and/or removed from a layer construction, and are then adiabatically deformed to generate another layer construction. Two layer constructions are regarded as topologically equivalent if they can be connected by such a bubbling process. Total inequivalent layer constructions yield the topological classification for the given symmetry class.

We first consider a $T$-invariant 3D insulator with glide symmetry $g_y$. In Fig.~\ref{fig-bubble}(a), for the $S_y$ spin bands, a pair of QSHI layers with opposite spin Chern numbers can emerge at $r_z=r_z^0$ from trivial spin bands or perturbations. Because $g_y$ persists in each sector of spin bands, another pair of QSHI layers should appear simultaneously at $r_z=r_z^0+\frac{c}{2}$ [Fig.~\ref{fig-bubble}(b)]. Without breaking symmetries, the two pairs of QSHI layers can adiabatically evolve toward layer A and layer B. Consequently, a change of the total spin Chern number $\Delta C_{s,y}^A=-\Delta C_{s,y}^B=+2$ occurs, showing that the $\bS_y$ spin bands are $\mathbb{Z}_2$-classified with $C_{s,y}^A=-C_{s,y}^B\ \text{mod}\ 2$ are topologically equivalent.
Artificially, a $\mathbb{Z}_2$ invariant can be defined as $C^{AB}_{s,y}=\left(C_{s,y}^A-C_{s,y}^B\right)/2\ \text{mod}\ 2$, and $C_{s,y}^{AB}=1$ indicates a $T$-DAXI. Moreover, for a gTCI with $g_y$, $C_{s,y}^{AB}=1$ inherits directly from the bulk $\mathbb{Z}_2$-hourglass invariant $\delta_h=1$ ($\delta_h=1$ corresponds to stacking QSHIs at layer A and layer B).
%This implies that the nonsymmorphic $T$-DAXI is more stable --- a spin gap closure and reopening cannot destroy the $T$-DAXI phase in a gTCI, which is contrast to the $T$-DAXI phase in the inversion TCI that its stability requires no spin gap closure~\cite{the-nc}.

For the $s_{xz}$ spin resolution, the bubble equivalence shown in Fig.~\ref{fig-bubble}(c,~d) demonstrates that adiabatic, symmetry-preserving perturbations do not alter the spin Chern numbers at layer A or B due to $g_yT$. Consequently, a $\mathbb{Z}$-classified 3D spin Chern number can be defined as $\gamma_{s,x/z}=C_{s,x/z}^A+C_{s,x/z}^B=2C_{s,x/z}^A$, which corresponds to a 3D QSHI with a topological contribution to the bulk SHC $e\gamma_{s,x/z}/2\pi$. The bubble-equivalence analysis for screw symmetry $s_2$ is presented in Fig.~\ref{fig-bubble-screw}, and we summarize the spin-resolved topological classifications discussed here in Table.~\ref{table: sLC classification} in the main text.

%\paragraph*{Spin-resolved topology and $T$-DAXI.}
\section{Spin-resolved nested Wilson loop characterization}
\label{SM: sNWL}
\begin{figure}[!t]
\centering
\includegraphics[width=1.03\linewidth]{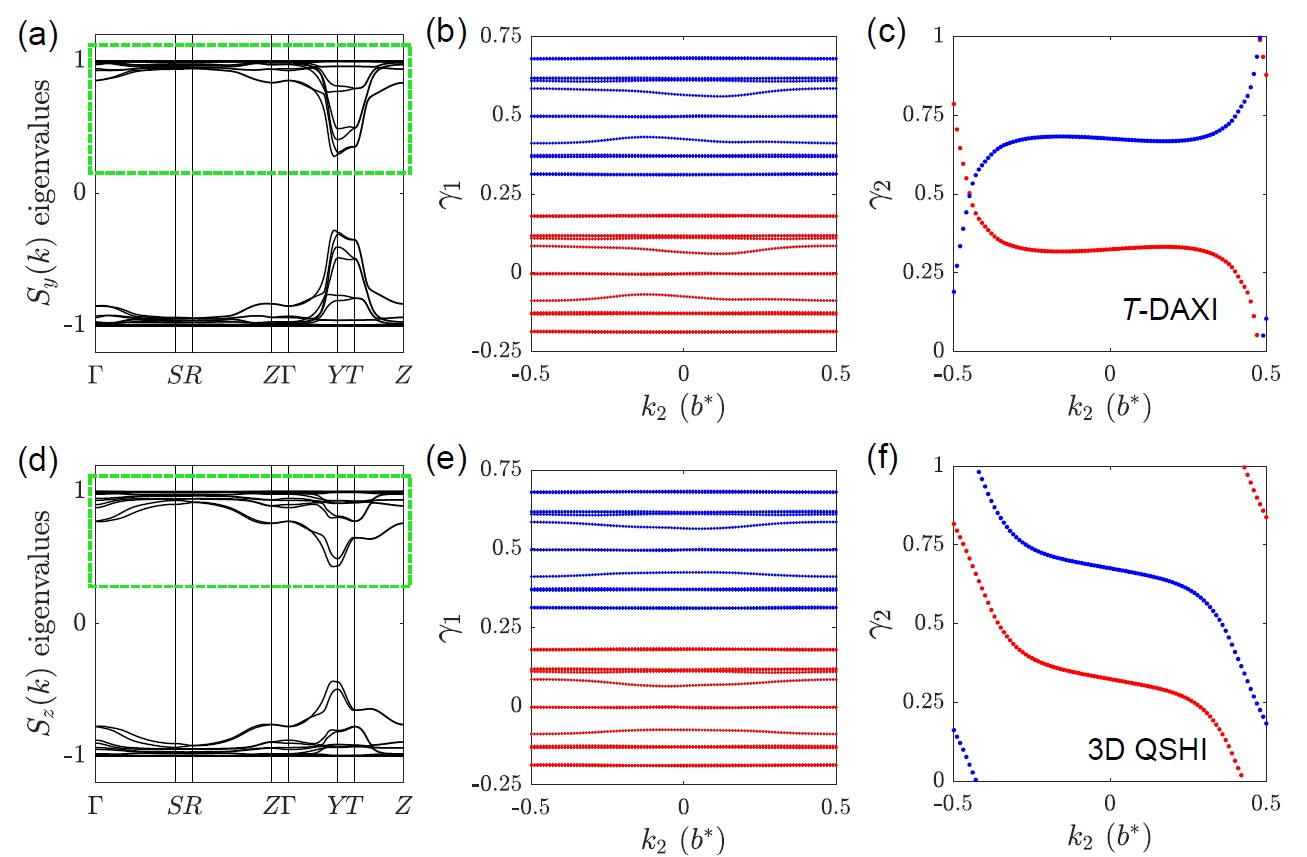}
\caption{(Color online) 
(a--c) Workflow of the spin-resolved nested Wilson loops in the $s_y$ resolution.
(a) Spin bands formed by eigenvalues of $\bS_y(\bk)$, where a spin-band gap is clearly observed.
(b) Partial WCCs $\gamma_1$ of the $k_z$-directed spin-resolved Wilson loops for the upper spin bands $\lambda^{+}$ [inside the dashed-green frame in (a)].
(c) The red (blue) dotted phases $\gamma_2$ of the $k_1$-directed spin-resolved nested Wilson loops calculated using the red (blue) dotted Wilson bands in (b).
%The opposite windings indicate a spin-resolved topology by placing 2D layers on the $(001;0)$ and $(001;\frac{1}{2})$ planes with partial Chern numbers $C^{+}=1$ and $C^{+}=-1$, respectively [see Fig.~\ref{fig-LC}(b)]. With time-reversal symmetry, this confirms a $T$-DAXI protected by $g_y$ and $s_2T$.
(d--f) Workflow for the spin-resolved 3D QSHI phase in the $s_z$ resolution with a 3D spin Chern number $C_s=2$.
} \label{fig-sNWL}
\end{figure}

In the spin-resolved topology, a $T$-DAXI can be considered as a stack of QSHI layers with alternating spin Chern numbers. This picture can be diagnosed by the spin-resolved nested Wilson loop method~\cite{the-nc}.
The $s_y$ spin bands of Bi$_4$Br$_3$I are shown in Fig.~\ref{fig-sNWL}(a). One can clearly see that the occupied bands are separated into upper (positive eigenvalues) and lower (negative eigenvalues) spin bands. For the upper spin bands, we compute the $k_z$-directed Wilson loops, and their phases (Wilson bands) are plotted in Fig.~\ref{fig-sNWL}(b).
The Wilson bands form two separate groups. One group (colored red) is located near $z=0$ and the other group (colored blue) is located near $z=\frac{1}{2}\bc$. We then compute the $k_1$-directed nested Wilson loops for each group of these Wilson bands, and the obtained nested Wilson-loop phases are shown in Fig.~\ref{fig-sNWL}(c). The opposite windings indicate that the red-colored states near $z=0$ have partial Chern number $C^+=+1$, while the blue-colored states near $z=\frac{1}{2}\bc$ have $C^+=-1$. Therefore, the corresponding opposite spin Chern numbers on adjacent layers signal a $T$-DAXI in the presence of $g_y$/$s_2T$ and time-reversal symmetry. This numerical result exactly matches the spin-resolved layer construction in Fig.~\ref{fig-LC}(b,d).
In Fig.~\ref{fig-sNWL}(d--f), we show another workflow of the spin-resolved nested Wilson loop for the $s_z$ resolution. This workflow shows that the upper spin sectors of both layers located near $z=0$ and $z=\frac{1}{2}\bc$ have the same partial Chern number $C^+=1$, indicating a 3D QSHI phase with a total 3D spin Chern number $C_s=2$.
These results agree with the spin-resolved topological classification in nonsymmorphic TCIs discussed in Section \ref{section: spin-resolved topology}.

\begin{figure}[!t]
\centering
\includegraphics[width=1\linewidth]{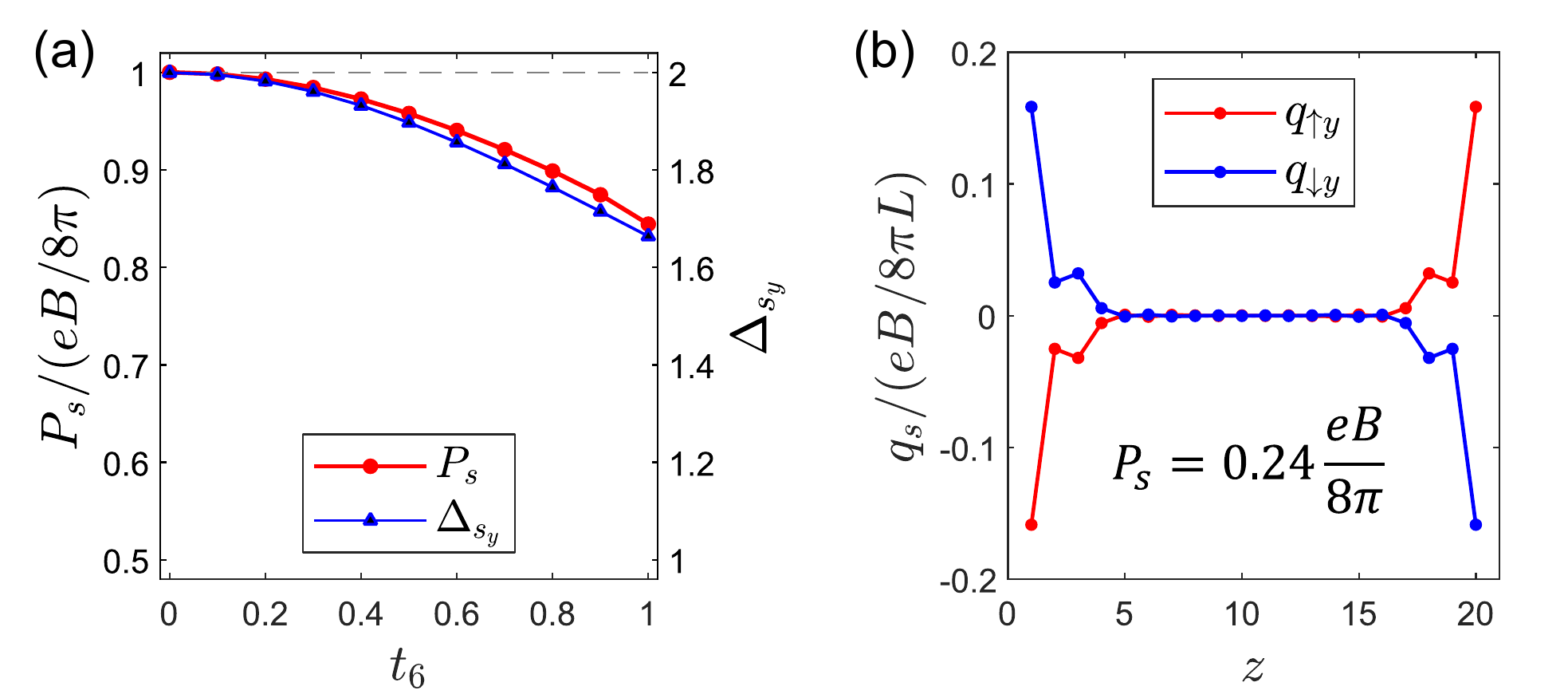}
\caption{(Color online) 
(a) The evolutions of the nonlocal spin polarization $P_s$ and the $s_y$-minimum spin gap $\Delta_{s_y}$ against the increasing of $s_y$-symmetry breaking strength. 
Modified parameters are $v_x=A_y=0$, $t_1=t_4=1.5$ and $-t_3=t_6$, with $t_6$ tuned from 0 to 1.
(b) The $s_y$ spin distributions in each unit cell of a 20-unit-cell slab under a $z$-directed magnetic field. Red and blue dotted lines indicate $s_y$ up and $s_y$ down, respectively. The total nonlocal spin polarization is $P_{s}=0.24eB/8\pi$.
Modified parameters are $t_1=t_4=0.4$ and $v_x=A_y=-t_3=t_6=2$, which give rise to a minimum $s_y$-spin gap $\Delta_{s_y}=0.36$.
} \label{Sfig-spin polarization}
\end{figure}

\section{Spin-magnetoelectric effect in spin non-conserving nonsymmorphic $T$-DAXIs}
\label{SM: spin-magnetoelectric-effect}
To demonstrate the spin-magnetoelectric effect in spin non-conserving nonsymmorphic $T$-DAXIs, we calculate real-space the nonlocal spin polarization using the gTCI model Eq.~\pref{Hamiltonian}, with parameters breaking the $s_y$ symmetry. Specifically, we tune the $s_y$-breaking terms $-t_3=t_6$ from 0 to 1. The evolutions of the resulting minimum spin gap of the $s_y$ spin spectrum $\Delta_{s_y}$ and the nonlocal spin polarization $P_s$ are shown in Fig.~\ref{Sfig-spin polarization}(a), demonstrated by the blue and red dotted lines, respectively.
One can see that when the spin conservation is relaxed, the nonlocal spin polarization perturbatively deviates from its quantized value, but lies close to that value if the spin-nonconserving terms are relatively weak.
We further decrease the minimum spin gap to $\Delta_{s_y}=0.36$, which is on the order of that in Bi$_4$Br$_3$I, and specifically calculate the $s_y$-spin distributions shown in Fig.~\ref{Sfig-spin polarization}(b). Opposite spins still clearly accumulate on opposite surfaces, resulting in a nonzero nonlocal spin polarization $P_s=0.24eB/8\pi$. 
Hence, we suggest that the nonsymmorphic real-material $T$-DAXI Bi$_4$Br$_3$I provides a promising platform for experimental detection of the spin-magnetoelectric effect.

\section{Calculation and methodology}
\label{SM: DFT-method}
We carried out first-principles calculations based on density functional theory (DFT) with the projector augmented wave (PAW) method~\cite{paw1, paw2}, as implemented in the Vienna \emph{ab initio} simulation package (VASP)~\cite{KRESSE199615, vasp}. The generalized gradient approximation (GGA) in the form of the Perdew-Burke-Ernzerhof (PBE) functional~\cite{pbe} was employed for the exchange-correlation potential. The kinetic energy cutoff for the plane-wave expansion was set to 450 eV. The Brillouin zone was sampled by the Monkhorst-Pack method in the self-consistent process, with a 16$\times$16$\times$4 $\textbf{k}$-mesh for the primitive cell.
The maximally localized Wannier functions and the Wannier-based tight-binding Hamiltonian for the Bi $p$, Br $p$ and I $p$ orbitals of Bi$_4$Br$_3$I were constructed using the Wannier90 package~\cite{wannier90-MOSTOFI20142309,wannier90}.
The (spin-resolved) nested Wilson loop of Bi$_4$Br$_3$I were calculated using the Wannier-based Hamiltonian.
The atomic coordinates and lattice constants of Bi$_4$Br$_3$I ($a=4.35585$ \AA, $b=13.4532$ \AA\ and $c=19.3272$ \AA) are adopted from Ref.~\cite{BiBrI-exp}. In our calculations of the mixed bismuth monohalide phase, the Br atoms occupy three of the four $4a$ Wyckoff positions while the I atoms occupy the remaining one $4a$ Wyckoff position.

%\begin{table}[!b]
%    \centering
%    \caption{Atomic coordinates of Bi$_4$Br$_3$I. Lattice constant refers to Ref.~\cite{BiBrI-exp}.}
%    \label{tab:Bi4Br3I_atomic_coordinates}
%    \small
%    \renewcommand{\arraystretch}{1.25}
%    \setlength{\tabcolsep}{0pt}
%    \begin{tabular*}{\columnwidth}{@{\extracolsep{\fill}}lcccc@{}}
%        \hline\hline
%        Site & Wyckoff & $x$ & $y$ & $z$ \\
%        \hline
%        Bi1 & 4$a$ & 0.00000 & 0.55697 & 0.99817 \\
%        Bi2 & 4$a$ & 0.00000 & 0.21722 & 0.00893 \\
%        Bi3 & 4$a$ & 0.00000 & 0.52481 & 0.15608 \\
%        Bi4 & 4$a$ & 0.00000 & 0.74767 & 0.35126 \\
%        Br1 & 4$a$ & 0.00000 & 0.87983 & 0.12789 \\
%        Br2 & 4$a$ & 0.00000 & 0.16729 & 0.19197 \\
%        Br3 & 4$a$ & 0.00000 & 0.39110 & 0.31624 \\
%        I   & 4$a$ & 0.00000 & 0.10591 & 0.38907 \\
%        \hline\hline
%    \end{tabular*}
%\end{table}

\bibliography{refs}

@article{the-nc,
  title={Spin-resolved topology and partial axion angles in three-dimensional insulators},
  author={Lin, Kuan-Sen and Palumbo, Giandomenico and Guo, Zhaopeng and Hwang, Yoonseok and Blackburn, Jeremy and Shoemaker, Daniel P. and Mahmood, Fahad and Wang, Zhijun and Fiete, Gregory A. and Wieder, Benjamin J. and Bradlyn, Barry},
  journal={Nature Communications},
  volume={15},
  pages={550},
  year={2024},
  doi = {10.1038/s41467-024-44762-w},
  url = {https://doi.org/10.1038/s41467-024-44762-w}
}

@article{BHZ,
author = {B. Andrei Bernevig  and Taylor L. Hughes  and Shou-Cheng Zhang },
title = {Quantum Spin Hall Effect and Topological Phase Transition in {HgTe} Quantum Wells},
journal = {Science},
volume = {314},
number = {5806},
pages = {1757-1761},
year = {2006},
doi = {10.1126/science.1133734},
URL = {https://www.science.org/doi/abs/10.1126/science.1133734}
}

@article{3DTI,
  title = {Topological Insulators in Three Dimensions},
  author = {Fu, Liang and Kane, C. L. and Mele, E. J.},
  journal = {Phys. Rev. Lett.},
  volume = {98},
  issue = {10},
  pages = {106803},
  numpages = {4},
  year = {2007},
  month = {Mar},
  publisher = {American Physical Society},
  doi = {10.1103/PhysRevLett.98.106803},
  url = {https://link.aps.org/doi/10.1103/PhysRevLett.98.106803}
}

@misc{Wieder2018,
      title={The Axion Insulator as a Pump of Fragile Topology}, 
      author={Benjamin J. Wieder and B. Andrei Bernevig},
      year={2018},
      eprint={1810.02373},
      archivePrefix={arXiv},
      url={https://arxiv.org/abs/1810.02373}, 
}

@article{Wilczek-AXI,
  title = {Two applications of axion electrodynamics},
  author = {Wilczek, Frank},
  journal = {Phys. Rev. Lett.},
  volume = {58},
  issue = {18},
  pages = {1799--1802},
  numpages = {0},
  year = {1987},
  month = {May},
  publisher = {American Physical Society},
  doi = {10.1103/PhysRevLett.58.1799},
  url = {https://link.aps.org/doi/10.1103/PhysRevLett.58.1799}
}

@article{Essin-AXI,
  title = {Magnetoelectric Polarizability and Axion Electrodynamics in Crystalline Insulators},
  author = {Essin, Andrew M. and Moore, Joel E. and Vanderbilt, David},
  journal = {Phys. Rev. Lett.},
  volume = {102},
  issue = {14},
  pages = {146805},
  numpages = {4},
  year = {2009},
  month = {Apr},
  publisher = {American Physical Society},
  doi = {10.1103/PhysRevLett.102.146805},
  url = {https://link.aps.org/doi/10.1103/PhysRevLett.102.146805}
}

@article{Olsen-AXI,
  title = {Surface theorem for the Chern-Simons axion coupling},
  author = {Olsen, Thomas and Taherinejad, Maryam and Vanderbilt, David and Souza, Ivo},
  journal = {Phys. Rev. B},
  volume = {95},
  issue = {7},
  pages = {075137},
  numpages = {14},
  year = {2017},
  month = {Feb},
  publisher = {American Physical Society},
  doi = {10.1103/PhysRevB.95.075137},
  url = {https://link.aps.org/doi/10.1103/PhysRevB.95.075137}
}

@article{magnetic-tqc,
  title={Magnetic topological quantum chemistry},
  author={Elcoro, Luis and Wieder, Benjamin J. and Song, Zhida and Xu, Yuanfeng and Bradlyn, Barry and Bernevig, B. Andrei},
  journal={Nature Communications},
  volume={12},
  pages={5965},
  year={2021},
  doi = {10.1038/s41467-021-26241-8},
  url = {https://doi.org/10.1038/s41467-021-26241-8}
}

@article{Aris-PRX,
  title = {Topological Insulators from Group Cohomology},
  author = {Alexandradinata, A. and Wang, Zhijun and Bernevig, B. Andrei},
  journal = {Phys. Rev. X},
  volume = {6},
  issue = {2},
  pages = {021008},
  numpages = {38},
  year = {2016},
  month = {Apr},
  publisher = {American Physical Society},
  doi = {10.1103/PhysRevX.6.021008},
  url = {https://link.aps.org/doi/10.1103/PhysRevX.6.021008}
}

@article{hourglass-fermion,
  title = {Hourglass fermions},
  author = {Wang, Zhijun and Alexandradinata, A. and Cava, R. J. and Bernevig, B. Andrei},
  journal = {Nature},
  volume = {532},
  issue = {7598},
  pages = {189194},
  numpages = {5},
  year = {2016},
  month = {Apr},
  doi = {10.1038/nature17410},
  url = {https://doi.org/10.1038/nature17410}
}

@article{LC-Song,
author = {Song, Zhida  and Zhang, Tiantian  and Fang, Zhong  and Fang, Chen},
title = {Quantitative mappings between symmetry and topology in solids},
journal = {Nature Communications},
volume = {9},
pages = {3530},
year = {2018},
doi = {10.1038/s41467-018-06010-w},
URL = {https://doi.org/10.1038/s41467-018-06010-w},
}

@article{Prodan,
  title = {Robustness of the spin-Chern number},
  author = {Prodan, Emil},
  journal = {Phys. Rev. B},
  volume = {80},
  issue = {12},
  pages = {125327},
  numpages = {7},
  year = {2009},
  month = {Sep},
  publisher = {American Physical Society},
  doi = {10.1103/PhysRevB.80.125327},
  url = {https://link.aps.org/doi/10.1103/PhysRevB.80.125327}
}

@article{Kane-Mele-1,
  title = {Quantum Spin Hall Effect in Graphene},
  author = {Kane, C. L. and Mele, E. J.},
  journal = {Phys. Rev. Lett.},
  volume = {95},
  issue = {22},
  pages = {226801},
  numpages = {4},
  year = {2005},
  month = {Nov},
  publisher = {American Physical Society},
  doi = {10.1103/PhysRevLett.95.226801},
  url = {https://link.aps.org/doi/10.1103/PhysRevLett.95.226801}
}

@article{Kane-Mele-2,
  title = {${Z}_{2}$ Topological Order and the Quantum Spin Hall Effect},
  author = {Kane, C. L. and Mele, E. J.},
  journal = {Phys. Rev. Lett.},
  volume = {95},
  issue = {14},
  pages = {146802},
  numpages = {4},
  year = {2005},
  month = {Sep},
  publisher = {American Physical Society},
  doi = {10.1103/PhysRevLett.95.146802},
  url = {https://link.aps.org/doi/10.1103/PhysRevLett.95.146802}
}

@article{wannier90,
  title = {Maximally localized Wannier functions: Theory and applications},
  author = {Marzari, Nicola and Mostofi, Arash A. and Yates, Jonathan R. and Souza, Ivo and Vanderbilt, David},
  journal = {Rev. Mod. Phys.},
  volume = {84},
  issue = {4},
  pages = {1419--1475},
  numpages = {0},
  year = {2012},
  month = {Oct},
  publisher = {American Physical Society},
  doi = {10.1103/RevModPhys.84.1419},
  url = {https://link.aps.org/doi/10.1103/RevModPhys.84.1419}
}

@article{wannier90-MOSTOFI20142309,
title = {An updated version of wannier90: A tool for obtaining maximally-localised Wannier functions},
journal = {Computer Physics Communications},
volume = {185},
number = {8},
pages = {2309-2310},
year = {2014},
issn = {0010-4655},
doi = {https://doi.org/10.1016/j.cpc.2014.05.003},
url = {https://www.sciencedirect.com/science/article/pii/S001046551400157X},
author = {Arash A. Mostofi and Jonathan R. Yates and Giovanni Pizzi and Young-Su Lee and Ivo Souza and David Vanderbilt and Nicola Marzari}
}

@article{paw1,
  title = {Projector augmented-wave method},
  author = {Bl\"ochl, P. E.},
  journal = {Phys. Rev. B},
  volume = {50},
  issue = {24},
  pages = {17953--17979},
  numpages = {0},
  year = {1994},
  month = {Dec},
  publisher = {American Physical Society},
  doi = {10.1103/PhysRevB.50.17953},
  url = {https://link.aps.org/doi/10.1103/PhysRevB.50.17953}
}

@article{paw2,
  title = {From ultrasoft pseudopotentials to the projector augmented-wave method},
  author = {Kresse, G. and Joubert, D.},
  journal = {Phys. Rev. B},
  volume = {59},
  issue = {3},
  pages = {1758--1775},
  numpages = {0},
  year = {1999},
  month = {Jan},
  publisher = {American Physical Society},
  doi = {10.1103/PhysRevB.59.1758},
  url = {https://link.aps.org/doi/10.1103/PhysRevB.59.1758}
}

@article{KRESSE199615,
title = "Efficiency of ab-initio total energy calculations for metals and semiconductors using a plane-wave basis set",
journal = "Computational Materials Science",
volume = "6",
number = "1",
pages = "15 - 50",
year = "1996",
issn = "0927-0256",
doi = "https://doi.org/10.1016/0927-0256(96)00008-0",
url = "http://www.sciencedirect.com/science/article/pii/0927025696000080",
author = "G. Kresse and J. Furthmüller",
}

@article{vasp,
  title = {Efficient iterative schemes for ab initio total-energy calculations using a plane-wave basis set},
  author = {Kresse, G. and Furthm\"uller, J.},
  journal = {Phys. Rev. B},
  volume = {54},
  issue = {16},
  pages = {11169--11186},
  numpages = {0},
  year = {1996},
  month = {Oct},
  publisher = {American Physical Society},
  doi = {10.1103/PhysRevB.54.11169},
  url = {https://link.aps.org/doi/10.1103/PhysRevB.54.11169}
}

@article{pbe,
  title = {Generalized Gradient Approximation Made Simple},
  author = {Perdew, John P. and Burke, Kieron and Ernzerhof, Matthias},
  journal = {Phys. Rev. Lett.},
  volume = {77},
  issue = {18},
  pages = {3865--3868},
  numpages = {0},
  year = {1996},
  month = {Oct},
  publisher = {American Physical Society},
  doi = {10.1103/PhysRevLett.77.3865},
  url = {https://link.aps.org/doi/10.1103/PhysRevLett.77.3865}
}

@Article{exp-MBT,
title = {Experimental Realization of an Intrinsic Magnetic Topological Insulator},
journal = {Chin. Phys. Lett.},
volume = {36},
number = {7},
pages = {076801-076801},
year = {2019},
issn = {},
doi = {10.1088/0256-307X/36/7/076801},	
url = {http://cpl.iphy.ac.cn/en/article/doi/10.1088/0256-307X/36/7/076801},
author = {Yan Gong and Jingwen Guo and Jiaheng Li and Kejing Zhu and Menghan Liao and Xiaozhi Liu and Qinghua Zhang and Lin Gu and Lin Tang and Xiao Feng and Ding Zhang and Wei Li and Canli Song and Lili Wang and Pu Yu and Xi Chen and Yayu Wang and Hong Yao and Wenhui Duan and Yong Xu and Shou-Cheng Zhang and Xucun Ma and Qi-Kun Xue and Ke He}
}

@article{MBT-Duan,
author = {Jiaheng Li  and Yang Li  and Shiqiao Du  and Zun Wang  and Bing-Lin Gu  and Shou-Cheng Zhang  and Ke He  and Wenhui Duan  and Yong Xu },
title = {Intrinsic magnetic topological insulators in van der Waals layered {MnBi$_2$Te$_4$}-family materials},
journal = {Science Advances},
volume = {5},
number = {6},
pages = {eaaw5685},
year = {2019},
doi = {10.1126/sciadv.aaw5685},
URL = {https://www.science.org/doi/abs/10.1126/sciadv.aaw5685},
}

@article{MBT-Zhang,
  title = {Topological Axion States in the Magnetic Insulator \uppercase{M}n\uppercase{B}i$_{2}$\uppercase{T}e$_{4}$ with the Quantized Magnetoelectric Effect},
  author = {Zhang, Dongqin and Shi, Minji and Zhu, Tongshuai and Xing, Dingyu and Zhang, Haijun and Wang, Jing},
  journal = {Phys. Rev. Lett.},
  volume = {122},
  issue = {20},
  pages = {206401},
  numpages = {6},
  year = {2019},
  month = {May},
  publisher = {American Physical Society},
  doi = {10.1103/PhysRevLett.122.206401},
  url = {https://link.aps.org/doi/10.1103/PhysRevLett.122.206401}
}

@article{EuInAs,
  title = {Axion insulator, Weyl points, quantum anomalous Hall effect, and magnetic topological phase transition in {${\mathrm{Eu}}_{3}{\mathrm{In}}_{2}{\mathrm{As}}_{4}$}},
  author = {Yao, Jingyu and Zhang, Ruihan and Zhang, Sheng and Sheng, Haohao and Shi, Youguo and Fang, Zhong and Weng, Hongming and Wang, Zhijun},
  journal = {Phys. Rev. B},
  volume = {111},
  issue = {4},
  pages = {L041117},
  numpages = {8},
  year = {2025},
  month = {Jan},
  publisher = {American Physical Society},
  doi = {10.1103/PhysRevB.111.L041117},
  url = {https://link.aps.org/doi/10.1103/PhysRevB.111.L041117}
}

@article{Z4-PRX,
  title = {Symmetry Indicators and Anomalous Surface States of Topological Crystalline Insulators},
  author = {Khalaf, Eslam and Po, Hoi Chun and Vishwanath, Ashvin and Watanabe, Haruki},
  journal = {Phys. Rev. X},
  volume = {8},
  issue = {3},
  pages = {031070},
  numpages = {33},
  year = {2018},
  month = {Sep},
  publisher = {American Physical Society},
  doi = {10.1103/PhysRevX.8.031070},
  url = {https://link.aps.org/doi/10.1103/PhysRevX.8.031070}
}

@article{Z4-nc,
  title = {Symmetry-based indicators of band topology in the 230 space groups},
  author = {Po, Hoi Chun and Vishwanath, Ashvin and Watanabe, Haruki},
  journal = {Nature Communications},
  volume = {8},
  issue = {1},
  pages = {50},
  year = {2017},
  month = {Jun},
  doi = {10.1038/s41467-017-00133-2},
  url = {https://doi.org/10.1038/s41467-017-00133-2}
}

@article{Z4-prb,
  title = {Higher-order topological insulators and superconductors protected by inversion symmetry},
  author = {Khalaf, Eslam},
  journal = {Phys. Rev. B},
  volume = {97},
  issue = {20},
  pages = {205136},
  numpages = {13},
  year = {2018},
  month = {May},
  publisher = {American Physical Society},
  doi = {10.1103/PhysRevB.97.205136},
  url = {https://link.aps.org/doi/10.1103/PhysRevB.97.205136}
}

@article{AXI-Qi-prb,
  title = {Topological field theory of time-reversal invariant insulators},
  author = {Qi, Xiao-Liang and Hughes, Taylor L. and Zhang, Shou-Cheng},
  journal = {Phys. Rev. B},
  volume = {78},
  issue = {19},
  pages = {195424},
  numpages = {43},
  year = {2008},
  month = {Nov},
  publisher = {American Physical Society},
  doi = {10.1103/PhysRevB.78.195424},
  url = {https://link.aps.org/doi/10.1103/PhysRevB.78.195424}
}

@article{dynamical-AXI-Qi-prb,
  title = {Dynamical axion field in topological magnetic insulators},
  author = {Li, Rundong and Wang, Jing and Qi, Xiao-Liang and Zhang, Shou-Cheng},
  journal = {Nature Physics},
  volume = {6},
  issue = {4},
  pages = {284},
  numpages = {5},
  year = {2010},
  month = {April},
  publisher = {Nature},
  doi = {10.1038/nphys1534},
  url = {https://doi.org/10.1038/nphys1534}
}

@article{MBT-exp-QAH,
  title = {Robust axion insulator and Chern insulator phases in a two-dimensional antiferromagnetic topological insulator},
  author = {Liu, Chang and Wang, Yongchao and Li, Hao and Wu, Yang and Li, Yaoxin and Li, Jiaheng and He, Ke and Xu, Yong and Zhang, Jinsong and Wang, Yayu},
  journal = {Nature Materials},
  volume = {19},
  issue = {5},
  pages = {522},
  numpages = {6},
  year = {2020},
  month = {May},
  publisher = {Nature},
  doi = {10.1038/s41563-019-0573-3},
  url = {https://doi.org/10.1038/s41563-019-0573-3}
}

@article{10.1093/nsr/nwad025,
    author = {Gong, Ming and Liu, Haiwen and Jiang, Hua and Chen, Chui-Zhen and Xie, X-C},
    title = {Half-quantized helical hinge currents in axion insulators},
    journal = {National Science Review},
    volume = {10},
    number = {9},
    pages = {nwad025},
    year = {2023},
    month = {02},
    issn = {2095-5138},
    doi = {10.1093/nsr/nwad025},
    url = {https://doi.org/10.1093/nsr/nwad025},
}

@article{PhysRevLett.129.096601,
  title = {Transport Theory of Half-Quantized Hall Conductance in a Semimagnetic Topological Insulator},
  author = {Zhou, Humian and Li, Hailong and Xu, Dong-Hui and Chen, Chui-Zhen and Sun, Qing-Feng and Xie, X. C.},
  journal = {Phys. Rev. Lett.},
  volume = {129},
  issue = {9},
  pages = {096601},
  numpages = {6},
  year = {2022},
  month = {Aug},
  publisher = {American Physical Society},
  doi = {10.1103/PhysRevLett.129.096601},
  url = {https://link.aps.org/doi/10.1103/PhysRevLett.129.096601}
}

@article{statistical-AXI-Song,
  title = {Delocalization Transition of a Disordered Axion Insulator},
  author = {Song, Zhi-Da and Lian, Biao and Queiroz, Raquel and Ilan, Roni and Bernevig, B. Andrei and Stern, Ady},
  journal = {Phys. Rev. Lett.},
  volume = {127},
  issue = {1},
  pages = {016602},
  numpages = {7},
  year = {2021},
  month = {Jun},
  publisher = {American Physical Society},
  doi = {10.1103/PhysRevLett.127.016602},
  url = {https://link.aps.org/doi/10.1103/PhysRevLett.127.016602}
}

@article{HOTI-MoTe2-Wang,
  title = {Higher-Order Topology, Monopole Nodal Lines, and the Origin of Large Fermi Arcs in Transition Metal Dichalcogenides $X{\mathrm{Te}}_{2}$ ($X=\mathrm{Mo},\mathrm{W}$)},
  author = {Wang, Zhijun and Wieder, Benjamin J. and Li, Jian and Yan, Binghai and Bernevig, B. Andrei},
  journal = {Phys. Rev. Lett.},
  volume = {123},
  issue = {18},
  pages = {186401},
  numpages = {9},
  year = {2019},
  month = {Oct},
  publisher = {American Physical Society},
  doi = {10.1103/PhysRevLett.123.186401},
  url = {https://link.aps.org/doi/10.1103/PhysRevLett.123.186401}
}

@article{axionangle-WCC-2020,
  title = {Axion coupling in the hybrid Wannier representation},
  author = {Varnava, Nicodemos and Souza, Ivo and Vanderbilt, David},
  journal = {Phys. Rev. B},
  volume = {101},
  issue = {15},
  pages = {155130},
  numpages = {26},
  year = {2020},
  month = {Apr},
  publisher = {American Physical Society},
  doi = {10.1103/PhysRevB.101.155130},
  url = {https://link.aps.org/doi/10.1103/PhysRevB.101.155130}
}

@article{AXI-Vanderbilt-2018,
  title = {Surfaces of axion insulators},
  author = {Varnava, Nicodemos and Vanderbilt, David},
  journal = {Phys. Rev. B},
  volume = {98},
  issue = {24},
  pages = {245117},
  numpages = {12},
  year = {2018},
  month = {Dec},
  publisher = {American Physical Society},
  doi = {10.1103/PhysRevB.98.245117},
  url = {https://link.aps.org/doi/10.1103/PhysRevB.98.245117}
}

@article{non-Abelian-Berry,
  title = {Orbital magnetization in crystalline solids: Multi-band insulators, Chern insulators, and metals},
  author = {Ceresoli, Davide and Thonhauser, T. and Vanderbilt, David and Resta, R.},
  journal = {Phys. Rev. B},
  volume = {74},
  issue = {2},
  pages = {024408},
  numpages = {13},
  year = {2006},
  month = {Jul},
  publisher = {American Physical Society},
  doi = {10.1103/PhysRevB.74.024408},
  url = {https://link.aps.org/doi/10.1103/PhysRevB.74.024408}
}

@article{axionterm-prl,
  title = {Adiabatic Pumping of Chern-Simons Axion Coupling},
  author = {Taherinejad, Maryam and Vanderbilt, David},
  journal = {Phys. Rev. Lett.},
  volume = {114},
  issue = {9},
  pages = {096401},
  numpages = {5},
  year = {2015},
  month = {Mar},
  publisher = {American Physical Society},
  doi = {10.1103/PhysRevLett.114.096401},
  url = {https://link.aps.org/doi/10.1103/PhysRevLett.114.096401}
}

@article{zoperator-1,
  title = {Maximally localized generalized Wannier functions for composite energy bands},
  author = {Marzari, Nicola and Vanderbilt, David},
  journal = {Phys. Rev. B},
  volume = {56},
  issue = {20},
  pages = {12847--12865},
  numpages = {0},
  year = {1997},
  month = {Nov},
  publisher = {American Physical Society},
  doi = {10.1103/PhysRevB.56.12847},
  url = {https://link.aps.org/doi/10.1103/PhysRevB.56.12847}
}

@article{zoperator-2,
  title = {Wannier-Based Definition of Layer Polarizations in Perovskite Superlattices},
  author = {Wu, Xifan and Di\'eguez, Oswaldo and Rabe, Karin M. and Vanderbilt, David},
  journal = {Phys. Rev. Lett.},
  volume = {97},
  issue = {10},
  pages = {107602},
  numpages = {4},
  year = {2006},
  month = {Sep},
  publisher = {American Physical Society},
  doi = {10.1103/PhysRevLett.97.107602},
  url = {https://link.aps.org/doi/10.1103/PhysRevLett.97.107602}
}

@article{review-JoAP,
    author = {Sekine, Akihiko and Nomura, Kentaro},
    title = {Axion electrodynamics in topological materials},
    journal = {Journal of Applied Physics},
    volume = {129},
    number = {14},
    pages = {141101},
    year = {2021},
    month = {04},
    issn = {0021-8979},
    doi = {10.1063/5.0038804},
    url = {https://doi.org/10.1063/5.0038804}
}

@article{EuSn2P2-exp,
author = {Gian Marco Pierantozzi  and Alessandro De Vita  and Chiara Bigi  and Xin Gui  and Hung-Ju Tien  and Debashis Mondal  and Federico Mazzola  and Jun Fujii  and Ivana Vobornik  and Giovanni Vinai  and Alessandro Sala  and Cristina Africh  and Tien-Lin Lee  and Giorgio Rossi  and Tay-Rong Chang  and Weiwei Xie  and Robert J. Cava  and Giancarlo Panaccione },
title = {Evidence of magnetism-induced topological protection in the axion insulator candidate EuSn$_2$P$_2$},
journal = {Proceedings of the National Academy of Sciences},
volume = {119},
number = {4},
pages = {e2116575119},
year = {2022},
doi = {10.1073/pnas.2116575119},
URL = {https://www.pnas.org/doi/abs/10.1073/pnas.2116575119}
}

@article{EuSn2P2-cal,
  author    = {Gui, Xin and Pletikosi{\'c}, Ivo and Cao, Huibo and Tien, Hung-Ju and Xu, Xitong and Zhong, Ruidan and Wang, Guangqiang and Chang, Tay-Rong and Jia, Shuang and Valla, Tonica and Xie, Weiwei and Cava, Robert J.},
  title     = {A New Magnetic Topological Quantum Material Candidate by Design},
  journal   = {ACS Central Science},
  year      = {2019},
  month     = {May},
  day       = {22},
  volume    = {5},
  number    = {5},
  pages     = {900--910},
  doi       = {10.1021/acscentsci.9b00202},
  url       = {https://doi.org/10.1021/acscentsci.9b00202},
  publisher = {American Chemical Society},
  issn      = {2374-7943}
}

@article{Cr-BiTe,
  author    = {Mogi, M. and Kawamura, M. and Yoshimi, R. and Tsukazaki, A. and Kozuka, Y. and Shirakawa, N. and Takahashi, K. S. and Kawasaki, M. and Tokura, Y.},
  title     = {A magnetic heterostructure of topological insulators as a candidate for an axion insulator},
  journal   = {Nature Materials},
  year      = {2017},
  month     = {May},
  day       = {1},
  volume    = {16},
  number    = {5},
  pages     = {516--521},
  doi       = {10.1038/nmat4855},
  url       = {https://doi.org/10.1038/nmat4855},
  issn      = {1476-4660}
}

@article{EuIn2As2-prl,
  title = {Higher-Order Topology of the Axion Insulator {${\mathrm{EuIn}}_{2}{\mathrm{As}}_{2}$}},
  author = {Xu, Yuanfeng and Song, Zhida and Wang, Zhijun and Weng, Hongming and Dai, Xi},
  journal = {Phys. Rev. Lett.},
  volume = {122},
  issue = {25},
  pages = {256402},
  numpages = {6},
  year = {2019},
  month = {Jun},
  publisher = {American Physical Society},
  doi = {10.1103/PhysRevLett.122.256402},
  url = {https://link.aps.org/doi/10.1103/PhysRevLett.122.256402}
}

@article{Cr-BiTe-scienceadvanced,
author = {Masataka Mogi  and Minoru Kawamura  and Atsushi Tsukazaki  and Ryutaro Yoshimi  and Kei S. Takahashi  and Masashi Kawasaki  and Yoshinori Tokura },
title = {Tailoring tricolor structure of magnetic topological insulator for robust axion insulator},
journal = {Science Advances},
volume = {3},
number = {10},
pages = {eaao1669},
year = {2017},
doi = {10.1126/sciadv.aao1669},
URL = {https://www.science.org/doi/abs/10.1126/sciadv.aao1669}
}

@article{Ta2Se8I,
  author    = {Gooth, J. and Bradlyn, B. and Honnali, S. and Schindler, C. and Kumar, N. and Noky, J. and Qi, Y. and Shekhar, C. and Sun, Y. and Wang, Z. and Bernevig, B. A. and Felser, C.},
  title     = {Axionic charge-density wave in the Weyl semimetal {(TaSe$_4$)$_2$I}},
  journal   = {Nature},
  year      = {2019},
  month     = {Nov},
  day       = {1},
  volume    = {575},
  number    = {7782},
  pages     = {315--319},
  doi       = {10.1038/s41586-019-1630-4},
  url       = {https://doi.org/10.1038/s41586-019-1630-4},
  issn      = {1476-4687}
}

@article{GdPtBi-2010,
  title = {Antiferromagnetic topological insulators},
  author = {Mong, Roger S. K. and Essin, Andrew M. and Moore, Joel E.},
  journal = {Phys. Rev. B},
  volume = {81},
  issue = {24},
  pages = {245209},
  numpages = {10},
  year = {2010},
  month = {Jun},
  publisher = {American Physical Society},
  doi = {10.1103/PhysRevB.81.245209},
  url = {https://link.aps.org/doi/10.1103/PhysRevB.81.245209}
}

@article{GdPtBi-2014,
  title = {Magnetic structure of {GdBiPt}: A candidate antiferromagnetic topological insulator},
  author = {M\"uller, R. A. and Lee-Hone, N. R. and Lapointe, L. and Ryan, D. H. and Pereg-Barnea, T. and Bianchi, A. D. and Mozharivskyj, Y. and Flacau, R.},
  journal = {Phys. Rev. B},
  volume = {90},
  issue = {4},
  pages = {041109},
  numpages = {5},
  year = {2014},
  month = {Jul},
  publisher = {American Physical Society},
  doi = {10.1103/PhysRevB.90.041109},
  url = {https://link.aps.org/doi/10.1103/PhysRevB.90.041109}
}

@article{review-naturereviewphysics,
  author    = {Nenno, Dennis M. and Garcia, Christina A. C. and Gooth, Johannes and Felser, Claudia and Narang, Prineha},
  title     = {Axion physics in condensed-matter systems},
  journal   = {Nature Reviews Physics},
  year      = {2020},
  month     = {Dec},
  volume    = {2},
  number    = {12},
  pages     = {682--696},
  doi       = {10.1038/s42254-020-0240-2},
  url       = {https://doi.org/10.1038/s42254-020-0240-2},
  issn      = {2522-5820}
}

@article{review-TI-2011,
  title = {Topological insulators and superconductors},
  author = {Qi, Xiao-Liang and Zhang, Shou-Cheng},
  journal = {Rev. Mod. Phys.},
  volume = {83},
  issue = {4},
  pages = {1057--1110},
  numpages = {0},
  year = {2011},
  month = {Oct},
  publisher = {American Physical Society},
  doi = {10.1103/RevModPhys.83.1057},
  url = {https://link.aps.org/doi/10.1103/RevModPhys.83.1057}
}

@article{review-TI-2010,
  title = {Colloquium: Topological insulators},
  author = {Hasan, M. Z. and Kane, C. L.},
  journal = {Rev. Mod. Phys.},
  volume = {82},
  issue = {4},
  pages = {3045--3067},
  numpages = {0},
  year = {2010},
  month = {Nov},
  publisher = {American Physical Society},
  doi = {10.1103/RevModPhys.82.3045},
  url = {https://link.aps.org/doi/10.1103/RevModPhys.82.3045}
}

@article{ACcurrent-2022,
  title = {Identifying axion insulator by quantized magnetoelectric effect in antiferromagnetic \uppercase{M}n\uppercase{B}i$_{2}$\uppercase{T}e$_{4}$ tunnel junction},
  author = {Li, Yu-Hang and Cheng, Ran},
  journal = {Phys. Rev. Res.},
  volume = {4},
  issue = {2},
  pages = {L022067},
  numpages = {6},
  year = {2022},
  month = {Jun},
  publisher = {American Physical Society},
  doi = {10.1103/PhysRevResearch.4.L022067},
  url = {https://link.aps.org/doi/10.1103/PhysRevResearch.4.L022067}
}

@article{AXI-AFM-2022,
    author = {Wan, Yuhao and Li, Jiayu and Liu, Qihang},
    title = {Topological magnetoelectric response in ferromagnetic axion insulators},
    journal = {National Science Review},
    volume = {11},
    number = {2},
    pages = {nwac138},
    year = {2022},
    month = {07},
    issn = {2095-5138},
    doi = {10.1093/nsr/nwac138},
    url = {https://doi.org/10.1093/nsr/nwac138}
}

@article{TME-2021,
  title = {Nonlocal sidewall response and deviation from exact quantization of the topological magnetoelectric effect in axion-insulator thin films},
  author = {Pournaghavi, N. and Pertsova, A. and MacDonald, A. H. and Canali, C. M.},
  journal = {Phys. Rev. B},
  volume = {104},
  issue = {20},
  pages = {L201102},
  numpages = {5},
  year = {2021},
  month = {Nov},
  publisher = {American Physical Society},
  doi = {10.1103/PhysRevB.104.L201102},
  url = {https://link.aps.org/doi/10.1103/PhysRevB.104.L201102}
}

@article{magnetic-tqc-Gao,
  title = {Magnetic band representations, {Fu-Kane-like} symmetry indicators, and magnetic topological materials},
  author = {Gao, Jiacheng and Guo, Zhaopeng and Weng, Hongming and Wang, Zhijun},
  journal = {Phys. Rev. B},
  volume = {106},
  issue = {3},
  pages = {035150},
  numpages = {6},
  year = {2022},
  month = {Jul},
  publisher = {American Physical Society},
  doi = {10.1103/PhysRevB.106.035150},
  url = {https://link.aps.org/doi/10.1103/PhysRevB.106.035150}
}

@article{magnetic-tqc-Xu,
  author    = {Xu, Yuanfeng and Elcoro, Luis and Song, Zhi-Da and Wieder, Benjamin J. and Vergniory, M. G. and Regnault, Nicolas and Chen, Yulin and Felser, Claudia and Bernevig, B. Andrei},
  title     = {High-throughput calculations of magnetic topological materials},
  journal   = {Nature},
  year      = {2020},
  month     = {Oct},
  day       = {1},
  volume    = {586},
  number    = {7831},
  pages     = {702--707},
  doi       = {10.1038/s41586-020-2837-0},
  url       = {https://doi.org/10.1038/s41586-020-2837-0},
  issn      = {1476-4687}
}

@article{TCI-2011,
  title = {Topological Crystalline Insulators},
  author = {Fu, Liang},
  journal = {Phys. Rev. Lett.},
  volume = {106},
  issue = {10},
  pages = {106802},
  numpages = {4},
  year = {2011},
  month = {Mar},
  publisher = {American Physical Society},
  doi = {10.1103/PhysRevLett.106.106802},
  url = {https://link.aps.org/doi/10.1103/PhysRevLett.106.106802}
}

@article{tqc-Wang,
  author    = {Vergniory, M. G. and Elcoro, L. and Felser, Claudia and Regnault, Nicolas and Bernevig, B. Andrei and Wang, Zhijun},
  title     = {A complete catalogue of high-quality topological materials},
  journal   = {Nature},
  year      = {2019},
  month     = {Feb},
  day       = {1},
  volume    = {566},
  number    = {7745},
  pages     = {480--485},
  doi       = {10.1038/s41586-019-0954-4},
  url       = {https://doi.org/10.1038/s41586-019-0954-4},
  issn      = {1476-4687}
}

@article{tqc-Fang,
  author    = {Zhang, Tiantian and Jiang, Yi and Song, Zhida and Huang, He and He, Yuqing and Fang, Zhong and Weng, Hongming and Fang, Chen},
  title     = {Catalogue of topological electronic materials},
  journal   = {Nature},
  year      = {2019},
  month     = {Feb},
  day       = {1},
  volume    = {566},
  number    = {7745},
  pages     = {475--479},
  doi       = {10.1038/s41586-019-0944-6},
  url       = {https://doi.org/10.1038/s41586-019-0944-6},
  issn      = {1476-4687}
}

@article{tqc-Wan,
  author    = {Tang, Feng and Po, Hoi Chun and Vishwanath, Ashvin and Wan, Xiangang},
  title     = {Comprehensive search for topological materials using symmetry indicators},
  journal   = {Nature},
  year      = {2019},
  month     = {Feb},
  day       = {1},
  volume    = {566},
  number    = {7745},
  pages     = {486--489},
  doi       = {10.1038/s41586-019-0937-5},
  url       = {https://doi.org/10.1038/s41586-019-0937-5},
  issn      = {1476-4687}
}

@article{bubble-equivalence-2020,
  author    = {Song, Zhida and Fang, Chen and Qi, Yang},
  title     = {Real-space recipes for general topological crystalline states},
  journal   = {Nature Communications},
  year      = {2020},
  month     = {Aug},
  day       = {21},
  volume    = {11},
  number    = {1},
  pages     = {4197},
  doi       = {10.1038/s41467-020-17685-5},
  url       = {https://doi.org/10.1038/s41467-020-17685-5},
  issn      = {2041-1723}
}

@article{general-spinChern-PRL,
  title = {Quantum Spin-Hall Effect and Topologically Invariant Chern Numbers},
  author = {Sheng, D. N. and Weng, Z. Y. and Sheng, L. and Haldane, F. D. M.},
  journal = {Phys. Rev. Lett.},
  volume = {97},
  issue = {3},
  pages = {036808},
  numpages = {4},
  year = {2006},
  month = {Jul},
  publisher = {American Physical Society},
  doi = {10.1103/PhysRevLett.97.036808},
  url = {https://link.aps.org/doi/10.1103/PhysRevLett.97.036808}
}

@article{general-spinChern-PRB,
  title = {Spin Hall conductivity in insulators with nonconserved spin},
  author = {Monaco, Domenico and Ul\ifmmode \check{c}\else \v{c}\fi{}akar, Lara},
  journal = {Phys. Rev. B},
  volume = {102},
  issue = {12},
  pages = {125138},
  numpages = {8},
  year = {2020},
  month = {Sep},
  publisher = {American Physical Society},
  doi = {10.1103/PhysRevB.102.125138},
  url = {https://link.aps.org/doi/10.1103/PhysRevB.102.125138}
}

@article{BiBrI-exp,
  author    = {Dikarev, E. V. and Popovkin, B. A. and Shevelkov, A. V.},
  title     = {New polymolecular bismuth monohalides: Synthesis and crystal structures of {Bi$_4$Br$_x$I$_{4-x}$} ($x$ = 1, 2, or 3)},
  journal   = {Russian Chemical Bulletin},
  year      = {2001},
  month     = {Dec},
  day       = {1},
  volume    = {50},
  number    = {12},
  pages     = {2304--2309},
  doi       = {10.1023/A:1015010907973},
  url       = {https://doi.org/10.1023/A:1015010907973},
  issn      = {1573-9171}
}

@article{BiBr-shh,
  title = {Three-dimensional topological ferroelectrics},
  author = {Sheng, Haohao and Zhang, Sheng and Fang, Zhong and Weng, Hongming and Wang, Zhijun},
  journal = {Phys. Rev. Lett.},
  pages = {},
  year = {2026},
  month = {Sep},
  publisher = {American Physical Society},
  doi = {10.1103/nwyq-3t1f},
  url = {https://link.aps.org/doi/10.1103/nwyq-3t1f}
}

@article{BBH,
  title = {Electric multipole moments, topological multipole moment pumping, and chiral hinge states in crystalline insulators},
  author = {Benalcazar, Wladimir A. and Bernevig, B. Andrei and Hughes, Taylor L.},
  journal = {Phys. Rev. B},
  volume = {96},
  issue = {24},
  pages = {245115},
  numpages = {59},
  year = {2017},
  month = {Dec},
  publisher = {American Physical Society},
  doi = {10.1103/PhysRevB.96.245115},
  url = {https://link.aps.org/doi/10.1103/PhysRevB.96.245115}
}

@article{HOTI-2017-PRL,
  title = {Reflection-Symmetric Second-Order Topological Insulators and Superconductors},
  author = {Langbehn, Josias and Peng, Yang and Trifunovic, Luka and von Oppen, Felix and Brouwer, Piet W.},
  journal = {Phys. Rev. Lett.},
  volume = {119},
  issue = {24},
  pages = {246401},
  numpages = {5},
  year = {2017},
  month = {Dec},
  publisher = {American Physical Society},
  doi = {10.1103/PhysRevLett.119.246401},
  url = {https://link.aps.org/doi/10.1103/PhysRevLett.119.246401}
}

@article{rotation-anomaly,
author = {Chen Fang  and Liang Fu },
title = {New classes of topological crystalline insulators having surface rotation anomaly},
journal = {Science Advances},
volume = {5},
number = {12},
pages = {eaat2374},
year = {2019},
doi = {10.1126/sciadv.aat2374},
URL = {https://www.science.org/doi/abs/10.1126/sciadv.aat2374},
}

@article{mirrorTCI,
  author    = {Hsieh, Timothy H. and Lin, Hsin and Liu, Junwei and Duan, Wenhui and Bansil, Arun and Fu, Liang},
  title     = {Topological crystalline insulators in the {SnTe} material class},
  journal   = {Nature Communications},
  year      = {2012},
  month     = {Jul},
  day       = {31},
  volume    = {3},
  number    = {1},
  pages     = {982},
  doi       = {10.1038/ncomms1969},
  url       = {https://doi.org/10.1038/ncomms1969},
  issn      = {2041-1723}
}

@article{BBH-science,
author = {Wladimir A. Benalcazar  and B. Andrei Bernevig  and Taylor L. Hughes },
title = {Quantized electric multipole insulators},
journal = {Science},
volume = {357},
number = {6346},
pages = {61-66},
year = {2017},
doi = {10.1126/science.aah6442},
URL = {https://www.science.org/doi/abs/10.1126/science.aah6442},
}

@article{d-2-Song,
  title = {$(d\ensuremath{-}2)$-Dimensional Edge States of Rotation Symmetry Protected Topological States},
  author = {Song, Zhida and Fang, Zhong and Fang, Chen},
  journal = {Phys. Rev. Lett.},
  volume = {119},
  issue = {24},
  pages = {246402},
  numpages = {5},
  year = {2017},
  month = {Dec},
  publisher = {American Physical Society},
  doi = {10.1103/PhysRevLett.119.246402},
  url = {https://link.aps.org/doi/10.1103/PhysRevLett.119.246402}
}

@article{HOTI-Schindler,
author = {Frank Schindler  and Ashley M. Cook  and Maia G. Vergniory  and Zhijun Wang  and Stuart S. P. Parkin  and B. Andrei Bernevig  and Titus Neupert },
title = {Higher-order topological insulators},
journal = {Science Advances},
volume = {4},
number = {6},
pages = {eaat0346},
year = {2018},
doi = {10.1126/sciadv.aat0346},
URL = {https://www.science.org/doi/abs/10.1126/sciadv.aat0346},
}

@article{HOTI-Ezawa,
  title = {Magnetic second-order topological insulators and semimetals},
  author = {Ezawa, Motohiko},
  journal = {Phys. Rev. B},
  volume = {97},
  issue = {15},
  pages = {155305},
  numpages = {5},
  year = {2018},
  month = {Apr},
  publisher = {American Physical Society},
  doi = {10.1103/PhysRevB.97.155305},
  url = {https://link.aps.org/doi/10.1103/PhysRevB.97.155305}
}

@article{Z4value-chiy,
author = {Benjamin J. Wieder  and Barry Bradlyn  and Zhijun Wang  and Jennifer Cano  and Youngkuk Kim  and Hyeong-Seok D. Kim  and Andrew M. Rappe  and C. L. Kane  and B. Andrei Bernevig },
title = {Wallpaper fermions and the nonsymmorphic Dirac insulator},
journal = {Science},
volume = {361},
number = {6399},
pages = {246-251},
year = {2018},
doi = {10.1126/science.aan2802},
URL = {https://www.science.org/doi/abs/10.1126/science.aan2802},
}

@article{spin-ME-2022,
  author  = {Schindler, Frank and Tsirkin, Stepan S. and Neupert, Titus and Bernevig, B. Andrei and Wieder, Benjamin J.},
  title   = {Topological zero-dimensional defect and flux states in three-dimensional insulators},
  journal = {Nature Communications},
  year    = {2022},
  volume  = {13},
  number  = {1},
  pages   = {5791},
  doi     = {10.1038/s41467-022-33471-x},
  url     = {https://doi.org/10.1038/s41467-022-33471-x},
  issn    = {2041-1723}
}

@article{spin-ME-2013,
  title = {Symmetry-protected topological phases with charge and spin symmetries: Response theory and dynamical gauge theory in two and three dimensions},
  author = {Ye, Peng and Wang, Juven},
  journal = {Phys. Rev. B},
  volume = {88},
  issue = {23},
  pages = {235109},
  numpages = {20},
  year = {2013},
  month = {Dec},
  publisher = {American Physical Society},
  doi = {10.1103/PhysRevB.88.235109},
  url = {https://link.aps.org/doi/10.1103/PhysRevB.88.235109}
}

@article{chiy-value-PRB,
  title = {Topology of nonsymmorphic crystalline insulators and superconductors},
  author = {Shiozaki, Ken and Sato, Masatoshi and Gomi, Kiyonori},
  journal = {Phys. Rev. B},
  volume = {93},
  issue = {19},
  pages = {195413},
  numpages = {28},
  year = {2016},
  month = {May},
  publisher = {American Physical Society},
  doi = {10.1103/PhysRevB.93.195413},
  url = {https://link.aps.org/doi/10.1103/PhysRevB.93.195413}
}

@article{Guo2022,
  author  = {Guo, Zhaopeng and Deng, Junze and Xie, Yue and Wang, Zhijun},
  title   = {Quadrupole topological insulators in {Ta$_2$M$_3$Te$_5$} ({M} = {Ni}, {Pd}) monolayers},
  journal = {npj Quantum Materials},
  year    = {2022},
  volume  = {7},
  number  = {1},
  pages   = {87},
  doi     = {10.1038/s41535-022-00498-8},
  url     = {https://doi.org/10.1038/s41535-022-00498-8},
  issn    = {2397-4648}
}

@article{PhysRevX.11.021016,
  title = {Time-Reversal Symmetry Breaking Driven Topological Phase Transition in {EuB$_{6}$}},
  author = {Gao, Shun-Ye and Xu, Sheng and Li, Hang and Yi, Chang-Jiang and Nie, Si-Min and Rao, Zhi-Cheng and Wang, Huan and Hu, Quan-Xin and Chen, Xue-Zhi and Fan, Wen-Hui and Huang, Jie-Rui and Huang, Yao-Bo and Pryds, Nini and Shi, Ming and Wang, Zhi-Jun and Shi, You-Guo and Xia, Tian-Long and Qian, Tian and Ding, Hong},
  journal = {Phys. Rev. X},
  volume = {11},
  issue = {2},
  pages = {021016},
  numpages = {11},
  year = {2021},
  month = {Apr},
  publisher = {American Physical Society},
  doi = {10.1103/PhysRevX.11.021016},
  url = {https://link.aps.org/doi/10.1103/PhysRevX.11.021016}
}

%\ \\

%\clearpage
%\newpage

%\begin{widetext}
%        \beginsupplement{}
%        \setcounter{section}{0}
%        %\renewcommand{\thesubsection}{\arabic{subsection}}
%        \renewcommand{\thesubsection}{\Alph{subsection}}
%        \renewcommand{\thesubsubsection}{\alph{subsubsection}}

%\section*{SUPPLEMENTARY MATERIAL}

%\end{widetext}

\end{document}